\documentclass[prd,eqsecnum,twocolumn,superscriptaddress,nofootinbib]{revtex4-1}

\usepackage{amsfonts}
\usepackage{mathtools}% Loads amsmath
\usepackage{amssymb}
\usepackage{amsthm}
\usepackage{bm}
\usepackage{dcolumn}
\usepackage{epsfig}
\usepackage{graphicx}
\usepackage{graphics}
\usepackage[latin1]{inputenc}
\usepackage{latexsym}
\usepackage{rotating}
\usepackage{xcolor}
\usepackage{gensymb}
\usepackage{hyperref}

\begin{document}

\title{Computing gravitational radiation from highly eccentric bound binaries\\in the small-mass-ratio limit}

\author{Diego Andr\'es Rivera Orona}
\affiliation{Department of Physics, MIT, Cambridge, MA 02139 USA}
%\affiliation{AOT Division, Los Alamos National Laboratory, Los Alamos, NM 87545 USA}

\author{Scott A.\ Hughes}
\affiliation{Department of Physics, MIT, Cambridge, MA 02139 USA}
\affiliation{MIT Kavli Institute, MIT, Cambridge, MA 02139 USA}

\begin{abstract}
Precise strong-gravity models describing the evolution of bound compact binary systems provide important information about the gravitational radiation such binaries generate and how they evolve due to the radiation's backreaction.  The limit of very high eccentricity, $e \to 1$, is challenging to model: frequency-domain techniques require many harmonics to accurately describe radiation in this limit, and time-domain techniques can be slow to converge due to the systems' long timescales.  Many recent astrophysical analyses have nonetheless emphasized the importance of understanding gravitational-wave emission in this regime, highlighting the fact that many systems may evolve through $e \simeq 1$ en route to becoming detectable sources, and may enter the sensitive band of detectors with substantial eccentricity.  In this paper, we show how to adapt frequency-domain methods for computing gravitational radiation in the small-mass-ratio limit to effectively handle high eccentricity.  The key piece of this analysis is an integral over the source of the frequency-domain Teukolsky equation.  We examine features of this integral that make it difficult to evaluate as $e \to 1$, and study methods to circumvent these difficulties.  We find a simple fix which allows us to compute gravitational-wave amplitudes and the associated fluxes of energy and angular momentum with ease up to eccentricities $e = 0.99$; with patience (and stamina), we have pushed to $e = 0.999$, though numerical accuracy appears to degrade for such extreme cases.  Interestingly, we find that the well-known and often used classic formulas of Peters and Mathews are subject to a systematic error which persists at quite wide separation.  This has implications for models of compact binary formation which evolve through  $e \simeq 1$.
\end{abstract}

\maketitle

\section{Introduction}

Gravitational wave (GW) emission tends to reduce a binary's eccentricity for most bound orbits.  The hypothesis of non-zero eccentricity for events measured by ground-based detectors has been tested, and some events may be consistent with small but non-zero eccentricity \cite{Romero-Shaw:2019itr, Wu:2020zwr, Lenon:2020oza, OShea:2021faf, Iglesias:2022xfc, Ramos-Buades:2023yhy, Bonino:2022hkj, Gamba:2021gap, Xu:2025ajj, Gupte:2026whi}.  The evidence, however, is sufficiently tentative that it is common for analyses to assume zero eccentricity \cite{LIGOScientific:2026ctl}.  It is nonetheless likely that many binaries had large eccentricity long before they evolved into the sensitive band of current GW antennae.

Recent work \cite{Qunbar:2023vys, Mancieri:2024sfy, Xuan:2025uvl} argues that dynamical interactions may drive some binaries to near unity eccentricity long before they enter the band of current detectors.  This may leave a ``tail'' of sources with substantial in-band eccentricity to be measured \cite{Xuan:2024ApJ}.  Future measurements using instruments with sensitivity at lower frequencies, especially the space-based Laser Interferometer Space Antenna (LISA), are expected to probe a much broader range of populations, many of which are likely to include sources with substantial eccentricity.  Though eccentricity complicates their signals, it also opens channels to learning about the astrophysical nature of these signals \cite{Xuan:2025PRD, Xuan:2026}.  Extreme mass-ratio inspirals (EMRIs), binaries consisting of a stellar mass ($1 - 100\,M_\odot$) secondary which inspirals through the strong field of a massive ($\sim 10^6\,M_\odot$) black hole, are expected to have substantial eccentricity when they radiate in the band of the LISA detector, and may evolve through epochs of very high eccentricity en route to becoming a LISA source \cite{Rom:2024nso, Mancieri:2025cmx}.  Such sources may be quite common \cite{Naoz:2023ApJ}, and LISA must be prepared to measure them with large eccentricity.  As pipelines and techniques are developed to measure such sources, it is now time to develop models which accurately characterize these sources and their GWs.

Focusing on small mass ratio makes it possible to model binary systems in general relativity using black hole perturbation theory (BHPT): one can treat the binary using the exact spacetime describing a black hole plus a small curvature perturbation arising from an orbiting body.  Over the past two decades, an extensive body of work and codes based on BHPT has been developed to study this limit \cite{Hughes:1999bq, Sago:2005fn, vandeMeent:2017bcc, Pound:2021qin, Fujita:2020zxe, Hughes:2021exa, Isoyama:2021jjd}.  Recent rapid progress on this problem \cite{Chua:2020stf, Katz:2021yft, Speri:2023jte, Chapman-Bird:2025xtd} has largely been driven by the need to develop science and data analysis infrastructure for future LISA observations \cite{Babak:2017tow, Berry:2019wgg}.  Most of this development has focused on systems with eccentricity $e \le 0.9$.  Several works looked at high eccentricity in the past, including foundational analyses relatively early in the history of BHPT studies \cite{Oohara:1984ck, Kojima:1984cc}, as well as some early applications of modern time-domain techniques \cite{Martel:2003jj}.  More recent work has examined the use of frequency-domain methods for unbound systems \cite{Hopper:2017iyq, Whittall:2023xjp}, and several very recent studies have used a variety of techniques to find high-precision solutions describing unbound or marginally bound orbits \cite{KhalvatiCapra, Warburton:2025ymy}.  Though there has been progress and understanding of the problem in this regime, mass production of BHPT solutions for $e$ close to unity has not yet been achieved.

The core difficulty which existing frequency-domain methods face at high eccentricity is technical.  Describing a high eccentricity system in the frequency domain requires computing a very large number of harmonics of the frequency $\Omega_r$ associated with radial orbital motion.  Especially as the radial mode number $n$ grows, finding a solution to the equations of BHPT involves integrating a function which executes approximately $n$ oscillations, modulated by a more slowly changing envelope that can vary by orders of magnitude over its domain.  Even if the amplitude of the integrand is large, the integral tends to be rather small because of the many oscillations.  Achieving good numerical accuracy in such circumstances can be challenging.

In this paper, we carefully examine what limits numerical accuracy when computing solutions to the equations of BHPT for high-eccentricity sources, and study methods to maintain good quality solutions.  We begin in Sec.\ \ref{sec:synopsis} with an overview of bound Kerr geodesic orbits, the GWs and tidal coupling produced by such orbits, and how an orbit's properties evolve due to the backreaction of GW emission and tidal coupling to the black hole.  This material has appeared in detail elsewhere, so we keep this discussion brief, focusing on presenting enough background to illustrate why ``standard'' solutions tend to perform poorly at high eccentricity.  In Sec.\ \ref{sec:standardcalc}, we describe the general form of the results we find using these methods for most bound black hole orbits, and demonstrate that these techniques do not yield accurate solutions as eccentricity approaches unity.  We argue that the root difficulty underlying the breakdown of these methods is the numerical challenge of performing a particular integral for high mode number when $e$ is close to 1.

In Sec.\ \ref{sec:high_e_calc}, we examine two methods for evaluating this integral and reliably computing BHPT solutions in this challenging regime.  In Sec.\ \ref{sec:be_cool_or_be_cast_out}, we simply break the integral into subdivisions; the total integral is found by summing all contributions from each subdivided interval.  We find that this extremely straightforward method significantly improves our ability to compute solutions for high mode number and as $e$ approaches 1.  We also examine a more sophisticated method based on integrating by parts.  As we show in Sec.\ \ref{sec:byparts}, although this technique works, it does not offer much advantage over the simpler subdivided integral approach, and in fact is significantly slower.  It is possible that integrating by parts may be useful for pushing these methods farther; we leave this possibility to future work.

We present several interesting first applications of this method in Sec.\ \ref{sec:results}.  We show that we can reliably compute the radiation for a range of black hole orbits out to $e = 0.99$; other results indicate that we can go somewhat farther than this, perhaps to $e = 0.997$ or so, though the computing time needed grows significantly as $e$ approaches 1.  We find evidence that numerical noise may limit our ability to study orbits with $e \ge 0.999$, at least using double-precision arithmetic for all numerical analyses.  We compare our strong-field results with the famous leading-order Peters and Mathews \cite{Peters:1963ux, Peters:1964zz} backreaction results.  Interestingly, we find an important offset in certain quantities even for wide orbits, where one might imagine weak-field expressions to be valid.  We argue that this offset will lead to a secularly accumulating error in long inspirals, skewing important analyses, such as studies of the formation of high-eccentricity binary systems.  This motivates the use of tools based on more accurate approaches to backreaction.  Finally, we examine the total energy and angular momentum radiated per orbit along sequences of increasing eccentricity, finding good convergence as $e \to 1$.  Indeed, by fitting to the data we compute, we estimate the total energy and angular momentum radiated in parabolic encounters, finding very good agreement with past work \cite{Warburton:2025ymy} which specialized to $e \ge 1$.

Certain formulas which are important but lengthy and not needed in the main body of the paper are presented in Appendix \ref{app:lengthy}.  Throughout this paper, we use relativist's units in which $G = 1$, $c = 1$.

\section{Synopsis: Bound black hole orbits and black hole perturbation theory}
\label{sec:synopsis}

We begin with a short summary of the most relevant details describing bound black hole orbits, the GWs produced by these orbits, and backreaction from these GWs.  This material has been presented at length elsewhere \cite{Schmidt:2002qk, Drasco:2005kz, Fujita:2009bp, vandeMeent:2019cam, Hughes:2021exa}, so we refer the reader to those references for detailed discussion.  We aim to provide enough detail for this manuscript to be self contained.

\subsection{Bound Kerr geodesics}
\label{sec:geodesics}

We first review how we describe and parameterize bound orbits of Kerr black holes.  In this paper, we work in Boyer-Lindquist coordinates.  In these coordinates, the Kerr spacetime describing a black hole of mass $M$ and spin angular momentum $|{\bf S}| = aM$ is given by
\begin{align} 
    ds^2 = &-\left(1-\frac{2Mr}{\Sigma}\right)dt^2+\frac{\Sigma}{\Delta}\,dr^2+\Sigma\, d\theta^2
    \nonumber\\
    & +\left(r^2+a^2+\frac{2Ma^2r}{\Sigma}\sin^2\theta\right)\sin^2\theta\,  d\phi^2
    \nonumber\\
    & -\frac{4Mar}{\Sigma} \sin^2\theta \, dt \, d\phi\;,
    \label{eq:kerrmetric}
\end{align}
where
\begin{equation}
    \Sigma = r^2 + a^2 \cos^2 \theta\;,\quad\Delta = r^2 - 2Mr + a^2\;.
    \label{eq:SigmaDelta}
\end{equation}
Note that $\Delta = 0$ at $r = r_\pm = M \pm \sqrt{M^2 - a^2}$; $r = r_+$ gives the coordinate location of the event horizon.  The polar angle $\theta$ is measured from the black hole's spin axis.  The motion in this spacetime of a test body of mass $\mu$ is governed by the equations
\begin{align}
\left(\frac{dr}{d\lambda}\right)^2 &= [E(r^2 + a^2) - aL_z]^2 - \Delta[r^2 + (L_z - aE)^2 + Q]
\nonumber\\
&\equiv R(r)\;,
\label{eq:rdot}\\
\left(\frac{d\theta}{d\lambda}\right)^2 &= Q - \cot^2\theta L_z^2 - a^2\cos^2\theta(1 - E^2)
\nonumber\\
&\equiv \Theta(\theta)\;,
\label{eq:thdot}\\
\frac{d\phi}{d\lambda} &= \csc^2\theta L_z + \frac{2Mr a E}{\Delta} - \frac{a^2 L_z}{\Delta}
\nonumber\\
&\equiv \Phi(r,\theta)\;,
\label{eq:phdot}\\
\frac{dt}{d\lambda} &= E\left[\frac{(r^2 + a^2)^2}{\Delta} - a^2\sin^2\theta\right] - \frac{2Mra L_z}{\Delta}
\nonumber\\
&\equiv T(r, \theta)\;.
\label{eq:tdot}
\end{align}
We have introduced the integrals of motion energy $E$ (per unit $\mu$), axial angular momentum $L_z$ (per unit $\mu$), and Carter constant $Q$ (per unit $\mu^2$).  These quantities and the orbiting body's rest mass $\mu$ are conserved along any geodesic orbit.  The energy $E$ and angular momentum $L_z$ are related to the spacetime's timelike and axial Killing vectors, respectively; the Carter constant $Q$ is related to the Kerr metric's Killing tensor.  Heuristically, the Carter constant can be regarded as the magnitude squared of the non-axial angular momentum: $Q \simeq |{\bf L}_\perp|^2 = |{\bf L}|^2 - L_z^2$.  For Schwarzschild black holes, $a = 0$, this heuristic relationship is exact.

In Eqs.\ (\ref{eq:rdot})--(\ref{eq:tdot}), we use {\it Mino time} $\lambda$ as our independent parameter along orbits.  Writing $d/d\lambda = \Sigma\,d/d\tau$, where $\tau$ is proper time along a geodesic, puts these equations into familiar forms that appear in many textbooks (e.g., Eqs.\ (33.32a--d) of Ref.\ \cite{mtw}).  Mino time separates the radial and polar motions from each other, facilitating a frequency-domain strong-field analysis of bound orbital motion.  Note that the radial equation (\ref{eq:rdot}) depends only on the coordinate $r$ and the polar equation (\ref{eq:thdot}) depends only on $\theta$.  From this, it is a simple matter to show that for bound orbits, the motion is periodic in $\lambda$:
\begin{equation}
    r(\lambda) = r(\lambda + n\Lambda_r)\;, \quad \theta(\lambda) = \theta(\lambda + k\Lambda_\theta)\;,
\end{equation}
where $n$ and $k$ are both integers.  Simple to use formulas exist for the periods $\Lambda_r$ and $\Lambda_\theta$ \cite{Fujita:2009bp}; we define the associated frequencies by $\Upsilon_{r,\theta} = 2\pi/\Lambda_{r,\theta}$.

The motions in $t$ and $\phi$ are the sum of secularly accumulating contributions and oscillatory functions:
\begin{align}
t(\lambda) &= t_0 + \Gamma\lambda + \Delta t[r(\lambda),\theta(\lambda)]\;,
\label{eq:t_of_lambda}
\\
\phi(\lambda) &= \phi_0 + \Upsilon_\phi\lambda + \Delta\phi[r(\lambda),\theta(\lambda)]\;.
\label{eq:phi_of_lambda}
\end{align}
In these equations, $t_0$ and $\phi_0$ set the values of $t$ and $\phi$ when $\lambda = 0$, and
\begin{equation}
\Gamma = \langle T(r, \theta)\rangle\rangle\;,\quad
\Upsilon_\phi = \langle \Phi(r,\theta)\rangle\;.
\label{eq:GammaUpsphi}
\end{equation}
The quantity $\Gamma$ describes the mean rate at which observer time $t$ accumulates per unit $\lambda$; the Mino-time frequency $\Upsilon_\phi$ describes the mean rate at which $\phi$ accumulates per unit $\lambda$.  The associated period is $\Lambda_\phi = 2\pi/\Upsilon_\phi$.  Simple formulas likewise exist for $\Gamma$ and $\Upsilon_\phi$ \cite{Fujita:2009bp}.  The average used in Eq.\ (\ref{eq:GammaUpsphi}) is given by
\begin{equation}
\langle f(r,\theta)\rangle = \frac{1}{\Lambda_r\Lambda_\theta}
\int_0^{\Lambda_r}d\lambda_r
\int_0^{\Lambda_\theta}d\lambda_\theta\,f[r(\lambda_r), \theta(\lambda_\theta)]\;.
\label{eq:rthetaaveraging}
\end{equation}
The functions $\Delta t(r, \theta)$ and $\Delta\phi(r,\theta)$ both average to zero under this operation; detailed forms for these expressions can be found in Refs.\ \cite{Drasco:2003ky, Fujita:2009bp}.  The ratio of the Mino-time frequencies to $\Gamma$ gives the observer-time frequencies:
\begin{equation}
\Omega_{r,\theta,\phi} = \Upsilon_{r,\theta,\phi}/\Gamma\;.
\end{equation}
These frequencies allow us to develop Fourier-domain descriptions of quantities sourced by black hole orbits \cite{Drasco:2003ky}.  From these quantities, we also deduce the periods associated with each of these coordinate motions:
\begin{equation}
    T_{r,\theta,\phi} = 2\pi/\Omega_{r,\theta,\phi}\;.
    \label{eq:perioddef}
\end{equation}
The period $T_r$ tells us how long it takes for an orbit to move through a complete cycle of its radial motion, and is particularly relevant for characterizing the eccentric orbits which are the focus of this analysis.

As described so far, orbits are described by their integrals of motion, the set $(E, L_z, Q)$.  For many purposes, it is more useful to describe them using their orbital geometry.  A bound black hole orbit oscillates over the range $r_{\rm min} \le r \le r_{\rm max}$, where the bounds $r_{\rm min/max}$ can be determined from Eq.\ (\ref{eq:rdot}) given $(E, L_z, Q)$ (they are the outermost roots at which $dr/d\lambda = 0$).  These turning points can be parameterized as
\begin{equation}
r_{\rm min/max} = \frac{p}{1 \pm e}\;,
\end{equation}
where $p$ is the orbit's semi-latus rectum, and $e$ its eccentricity.  When $p \gg M$, these parameters characterize the elliptical orbits that emerge in the Newtonian limit.

A bound black hole orbit likewise oscillates over the range $\theta_{\rm min} \le \theta \le \theta_{\rm max}$, where $\theta_{\rm max} = \pi - \theta_{\rm min}$, and $\theta_{\rm min}$ corresponds to the real root\footnote{Given $(E, L_z, Q)$, there is another root for which $|\cos\theta_{\rm min}| > 1$, which does not correspond to a physical turning point.} of Eq.\ (\ref{eq:thdot}).  It has been found to be quite useful to parameterize orbits by an inclination $I$ defined such that
\begin{equation}
I = \pi/2 - {\rm sgn}(L_z)\theta_{\rm min}\;.
\end{equation}
It is simple to show that $\cos\theta_{\rm min} = \sin I$.  The angle $I$ varies smoothly from 0 for equatorial prograde orbits to $\pi$ for equatorial retrograde orbits; the orbit integral $L_z$ has the same sign as $x_I \equiv \cos I$.

Using this parameterization, we rewrite the radial and polar motions as
\begin{align}
r &= \frac{p}{1 + e\cos(\chi_r + \chi_{r0})}\;,
\label{eq:rdef}\\
\cos\theta &= \sqrt{1 - x_I^2}\cos(\chi_\theta + \chi_{\theta0})\;.
\label{eq:thetadef}
\end{align}
This form introduces the radial and polar anomaly angles $\chi_r$ and $\chi_\theta$.  These angles grow monotonically; the orbit oscillates between the turning points as they grow.  We put $\chi_\theta = 0$, $\chi_r = 0$, $t = t_0$, and $\phi = \phi_0$ when $\lambda = 0$.  The phase $\chi_{r0}$ then determines the value of $r$ at $\lambda = 0$, and $\chi_{\theta0}$ determines the corresponding value of $\theta$.

\subsection{Computing gravitational waves and the on-horizon tide from black hole orbits}
\label{sec:amps_and_evolve}

We next briefly summarize how we compute the distant GWs and the on-horizon tidal field arising from these orbits.  We refer the reader to Refs.\ \cite{Drasco:2005kz, OSullivan:2014ywd} for detailed discussion; this text closely follows a similar synopsis presented in \cite{Hughes:2021exa}.  Our goal here is to set up a discussion of why this computation is challenging when $e \to 1$, and how we deal with this challenge.

We begin with the Teukolsky equation \cite{Teukolsky:1973ha}, which describes perturbations of various fields to Kerr black holes.  We focus on the Newman-Penrose curvature scalar
\begin{equation}
\psi_4 = -C_{\alpha\beta\gamma\delta}n^\alpha\bar m^\beta n^\gamma\bar m^\delta\
\end{equation}
In this quantity, $C_{\alpha\beta\gamma\delta}$ is the Weyl curvature tensor, and $n^\alpha$ and $\bar m^\alpha$ are legs of the Newman-Penrose null tetrad \cite{Newman:1961qr}.  The field $\psi_4$ has spin-weight $s = -2$, and corresponds to outgoing gravitational radiation far from the source:
\begin{equation}
\psi_4 = \frac{1}{2}\frac{d^2}{dt^2}\left(h_+ - ih_\times\right)\quad\mbox{as $r\to\infty$}\;.
\label{eq:psi4_to_h}
\end{equation}
As $r \to r_+$, $\psi_4$ encodes tidal interactions of the orbiting body with the black hole's event horizon.

Teukolsky showed that $\psi_4$ is governed by the equation
\begin{widetext}
\begin{align}
&
\left[\frac{(r^2 + a^2)^2 }{\Delta} - a^2\sin^2\vartheta\right]\partial^2_{t}\Psi - 4\left[r + ia\cos\vartheta - \frac{M(r^2 - a^2)}{\Delta}\right]\partial_t\Psi +\frac{4 M a r}{\Delta}\partial_\varphi\partial_t\Psi- \Delta^{2}\partial_r\left(\Delta^{-1}\partial_r\Psi\right) 
\nonumber\\
&
- \frac{1}{\sin\vartheta}\partial_\vartheta \left(\sin\vartheta\partial_\vartheta\Psi\right) + \left[\frac{a^2}{\Delta}   -\frac{1}{\sin^2\vartheta}\right]\partial_\varphi^2 \Psi + 4 \left[\frac{a (r - M)}{\Delta} + \frac{i \cos\vartheta}{\sin^2\vartheta} \right]\partial_\varphi\Psi + \left(4\cot^2\vartheta + 2\right) \Psi = 4\pi\Sigma{\mathcal T}\;.
\label{eq:teuk}
\end{align}
\end{widetext}
The field $\Psi = (r - ia\cos\vartheta)^4\psi_4$, and ${\mathcal T}$ is a source term whose form is discussed further below.  See Ref.\ \cite{Teukolsky:1973ha} for additional details and definitions.

The frequency-domain approach we use to solve Eq.\ (\ref{eq:teuk}) begins by writing $\psi_4$ in a Fourier and multipole expansion:
\begin{widetext}
\begin{equation}
\psi_4 = \frac{1}{(r - ia\cos\vartheta)^4}\int_{-\infty}^{\infty}d\omega\sum_{l = 2}^\infty\sum_{m = -l}^l R_{lm\omega}(r)S_{lm\omega}(\vartheta)e^{i[m\varphi - \omega t]}\;.
\label{eq:psi4decomp}
\end{equation}
\end{widetext}
The coordinates $(t,r,\vartheta,\varphi)$ describe the field point where $\psi_4$ is measured; note the distinction from the orbit's polar and axial angles, $\theta$ and $\phi$.  Using this decomposition, Eq.\ (\ref{eq:teuk}) separates, with ordinary differential equations governing the $r$ and $\vartheta$ dependence.  The angular function $S_{lm\omega}(\vartheta)$ is a spin-weight $-2$ spheroidal harmonic; this function and the methods we use to compute it are discussed at length in Appendix A of Ref.\ \cite{Hughes:1999bq}.  The radial function $R_{lm\omega}(r)$ satisfies the equation
\begin{equation}
    \Delta^2\frac{d}{dr}\left(\frac{dR_{lm\omega}}{dr}\right) - V_{lm\omega}(r)R_{lm\omega} = \mathcal{T}_{lm\omega}(r)\;.
    \label{eq:radialteuk}
\end{equation}
The detailed form of the potential $V_{lm\omega}(r)$ can be found in Refs.\ \cite{Teukolsky:1973ha, Drasco:2005kz}; these details are not needed for this paper's analysis.

We solve Eq.\ (\ref{eq:radialteuk}) by first finding its homogeneous solutions.  Setting $\mathcal{T}_{lm\omega}(r) = 0$, (\ref{eq:radialteuk}) admits two independent solutions.  The solution $R^{\rm up}_{lm\omega}(r)$ describes an outgoing plane wave as $r \to \infty$, and is a mix of ingoing and outgoing waves away from this asymptotic region.  The solution $R^{\rm in}_{lm\omega}(r)$ describes an ingoing plane wave as $r \to r_+$, and is likewise a mix away from this region.  From these homogeneous solutions, we then construct a Green's function which, when integrated against the source $\mathcal{T}_{lm\omega}(r)$, yields a particular solution for $R_{lm\omega}(r)$.

The source $\mathcal{T}_{lm\omega}(r)$ on the right-hand side of Eq.\ (\ref{eq:radialteuk}) is essentially a Fourier and multipole mode of the source on the right-hand side of Eq.\ (\ref{eq:teuk}).  Of key importance for us is that that this source is built from the stress-energy tensor of a body orbiting a Kerr black hole,
\begin{align}
    T_{\alpha\beta} &= \mu\int d\tau\,\delta^{(4)}\left[x^\gamma - z^\gamma(\tau)\right]
    \nonumber\\
    &= \frac{\mu\,u_\alpha u_\beta}{(dt/d\tau)\Sigma\sin \vartheta}\delta[r-r(t)]\delta[\vartheta-\theta(t)]\delta[\varphi-\phi(t)]\;,
    \label{eq:stressenergy}
\end{align}
where $x^\gamma$ is a spacetime field point, $z^\gamma(\tau)$ is the worldline of the orbiting body, $u^\alpha = dz^\alpha/d\tau$ is the 4-velocity along that worldline, and $\delta^{(4)}(x^\gamma)$ is a delta function normalized with the Kerr metric's proper volume element.  This tensor is contracted with certain ``legs'' of the Newman-Penrose null tetrad:
\begin{align}
    n^\alpha &\doteq \frac{1}{2\Sigma}\left[(r^2 + a^2),-\Delta,0,a\right]\;,
    \label{eq:nNP}\\
    \bar m^\alpha &\doteq \rho\left[-ia\sin\theta,0,1,-i\csc\theta\right]\;,
    \label{eq:mbarNP}
\end{align}
where $\rho = 1/(r - ia\cos\theta)$.  We discuss further aspects of this term (or of a function derived from it) in Sec.\ \ref{sec:source}.  A key point we emphasize now is that this source involves polynomial factors of the orbital radius $r$, as well as the radial 4-velocity component $dr/d\tau$.  These quantities can vary by very large factors as eccentricity $e \to 1$.

We are particularly interested in computing $\psi_4$ in the asymptotic regimes $r \to \infty$ and $r \to r_+ = M + \sqrt{M^2 - a^2}$, the coordinate location of the black hole's event horizon.  The far solution is simple:
\begin{equation}
    \psi_4 = \frac{1}{r}\sum_{lmkn} Z^{\infty}_{lmkn}S_{lm\omega_{mkn}}(\vartheta)e^{i(\omega_{mkn}t - m\varphi)}\quad r\to\infty\;,
    \label{eq:psi4sum}
\end{equation}
where
\begin{equation}
    \omega_{mkn} = m\Omega_\phi + k\Omega_\theta + n\Omega_r\;.
\end{equation}
The amplitude $Z^{\infty}_{lmkn}$ is computed by integrating over a term constructed from the source function $\mathcal{T}_{lm\omega}(r,\theta)$, and is discussed in more detail in Sec.\ \ref{sec:source}.  The sum over $l$ in Eq.\ (\ref{eq:psi4sum}) is from $2$ to $\infty$; the sum over $m$ is from $-l$ to $l$; the sums over $k$ and $n$ are from $-\infty$ to $\infty$.  In practice, the sums to infinity are truncated at a finite value by assessing numerical convergence.  The amplitudes also respect the following symmetry:
\begin{equation}
    Z^\infty_{l,-m,-k,-n} = (-1)^{(l+k)}{\bar Z}^\infty_{lmkn}\;.
    \label{eq:Zsymmetry}
\end{equation}
We take advantage of this by only directly computing amplitudes with $n \ge 0$; those with $n < 0$ are found using this symmetry.  With $Z^\infty_{lmkn}$ in hand, it is a simple matter to compute the GWs produced by a black hole orbit, and how the GWs backreact on the source orbit's integrals of motion:
\begin{equation}
    h_+ - ih_\times = -\frac{2}{r}\sum_{lmkn} \frac{Z^{\infty}_{lmkn}}{\omega_{mkn}^2}S_{lm\omega_{mkn}}(\vartheta)e^{i(\omega_{mkn}t - m\varphi)}\;,
\end{equation}
\begin{align}
    \left(\frac{dE}{dt}\right)^\infty &= \sum_{lmkn} \frac{|Z^{\infty}_{lmkn}|^2}{4\pi\omega_{mkn}^2}\;,
    \label{eq:dEdtinf}\\
    \left(\frac{dL_z}{dt}\right)^\infty &= \sum_{lmkn} \frac{m|Z^{\infty}_{lmkn}|^2}{4\pi\omega_{mkn}^3}\;,
    \label{eq:dLzdtinf}\\
    \left(\frac{dQ}{dt}\right)^\infty &= \sum_{lmkn} |Z^{\infty}_{lmkn}|^2\frac{({\mathcal L}_{mkn} + k\Upsilon_\theta)}{2\pi\omega_{mkn}^3}\;.
    \label{eq:dQdtinf}
\end{align}
The factor $\mathcal{L}_{mkn}$ is written out in Appendix \ref{app:lengthy}.  The result for $h_+$ and $h_\times$ comes from combining Eqs.\ (\ref{eq:psi4sum}) and (\ref{eq:psi4_to_h}); the rates of change of $E$ and $L_z$ are derived in Ref.\ \cite{Teukolsky:1973ha}; the rate of change of $Q$ is given in Ref.\ \cite{Sago:2005fn}.

Finding $\psi_4$ as $r \to r_+$ allows us to compute the tidal coupling of the orbiting body with the event horizon.  Details of this analysis can be found in Ref.\ \cite{OSullivan:2014ywd}.  For us, the relevant result is that this solution is built from a set of amplitudes $Z^{\rm H}_{lmkn}$ that are found using an integral quite similar to the one that yields $Z^\infty_{lmkn}$.  This term also respects the symmetry (\ref{eq:Zsymmetry}) (replacing the ``$\infty$'' label with ``${\rm H}$''), and is used in sums much like Eq.\ (\ref{eq:psi4sum}).  This tidal coupling backreacts on the orbit, leading to additional contributions to the secular evolution of the orbit integrals:
\begin{align}
    \left(\frac{dE}{dt}\right)^{\rm H} &= \sum_{lmkn} \frac{\alpha_{lmkn}|Z^{\rm H}_{lmkn}|^2}{4\pi\omega_{mkn}^2}\;,
    \label{eq:dEdthrz}\\
    \left(\frac{dL_z}{dt}\right)^{\rm H} &= \sum_{lmkn} \frac{\alpha_{lmkn}m|Z^{\rm H}_{lmkn}|^2}{4\pi\omega_{mkn}^3}\;,
    \label{eq:dLzdthrz}\\
    \left(\frac{dQ}{dt}\right)^{\rm H} &= \sum_{lmkn} \alpha_{lmkn}|Z^{\rm H}_{lmkn}|^2\frac{({\mathcal L}_{mkn} + k\Upsilon_\theta)}{2\pi\omega_{mkn}^3}\;.
    \label{eq:dQdthrz}
\end{align}
The (somewhat complicated) factor $\alpha_{lmkn}$ which appears in these expressions is written out in Appendix \ref{app:lengthy}. 

Note that Eqs.\ (\ref{eq:psi4_to_h}) and (\ref{eq:psi4sum}) have a hidden factor: the secondary's mass $\mu$. One should regard the quantities $\psi_4$ and $h$ to be per unit $\mu$.  This factor arises from the factor of $\mu$ in the stress-energy tensor of the orbiting body, as in Eq.\ (\ref{eq:stressenergy}).  Similarly, when the dimensions of various terms are fully taken into account, one finds that Eqs.\ (\ref{eq:dEdtinf}) and (\ref{eq:dEdthrz}) are dimensionless in relativist's units, but include hidden factors of mass-ratio squared, $(\mu/M)^2$; one should regard these quantities as per unit squared mass-ratio.  The angular momentum fluxes (\ref{eq:dLzdtinf}) and (\ref{eq:dLzdthrz}) have the dimension of mass, and should be regarded as per unit $(\mu^2/M)$.  Finally, the rates of change of Carter constant, (\ref{eq:dQdtinf}) and (\ref{eq:dQdthrz}), have the dimension of mass squared, and should be regarded as per unit $\mu^2$.  Keeping these ``hidden'' proportionalities in mind is particularly important when comparing our numerical results to weak-field analytic predictions.

Computing GWs and the on-horizon tide from a black hole orbit, and the backreaction of those quantities on the orbit, thus boils down to computing the amplitudes $Z^{\infty,{\rm H}}_{lmkn}$.  We now discuss how this is done, and then turn to a discussion of the challenge of evaluating these terms as $e \to 1$.

\subsection{Source integral}
\label{sec:source}

The amplitudes $Z^{\infty,{\rm H}}_{lmkn}$ are found by computing Fourier modes of a function $I^{\infty,{\rm H}}_{lm\omega}(r,\theta)$ which is in turn constructed from the radial Teukolsky equations source term (\ref{eq:radialteuk}) and its homogeneous solutions:
\begin{widetext}
\begin{equation}
    Z^{\infty,{\rm H}}_{lmkn} = \frac{e^{-im\phi_0}}{\Lambda_r\Lambda_\theta\Gamma}\int_0^{\Lambda_r}d\lambda_r\,e^{in\Upsilon_r\lambda_r} \int_0^{\Lambda_\theta}d\lambda_\theta\,e^{ik\Upsilon_\theta\lambda_\theta}\left(\frac{dt}{d\lambda}\right) I^{\infty,{\rm H}}_{lm\omega_{mkn}}(r,\theta)e^{i[\omega_{mkn}\Delta t(r,\theta) - m\Delta\phi(r,\theta)]}\;.
    \label{eq:Zlmkn_integral}
\end{equation}
Where $r$ appears in the integrand of Eq.\ (\ref{eq:Zlmkn_integral}), it should be read as $r(\lambda_r)$; likewise, $\theta$ in this integrand should be read as $\theta(\lambda_\theta)$.  Note that $dt/d\lambda$ is also a function of $r$ and $\theta$.  A derivation of the function $I^{\infty, H}_{lm\omega}(r,\theta)$ is presented in Refs.\ \cite{Drasco:2005kz, OSullivan:2014ywd}; it is given by
\begin{equation}
    I^{\infty,{\rm H}}_{lm\omega}(r,\theta) = \left[\left(A_{nn0}+A_{n\bar m 0} + A_{\bar m \bar m 0}\right) - \left(A_{n\bar m 1} + A_{\bar m \bar m 1}\right)\frac{d}{dr} + A_{\bar m \bar m 2}\frac{d^2}{dr^2} \right] R^{{\rm in},{\rm up}}_{lm\omega}(r)\;.
    \label{eq:source_integrand}
\end{equation}
When this function is used in Eq.\ (\ref{eq:Zlmkn_integral}), it is evaluated using the frequency harmonic $\omega = \omega_{mkn}$.  Note that the ``in'' homogeneous solution is used to compute $Z^\infty_{lmkn}$; the ``up'' solution is used to compute $Z^{\rm H}_{lmkn}$.  See Appendix D of \cite{OSullivan:2014ywd} for details and further discussion.
\end{widetext}

The terms $A_{abj}$ that appear in (\ref{eq:source_integrand}) are built by contracting the stress-energy tensor (\ref{eq:stressenergy}) with the Newman-Penrose tetrad legs $n^\alpha$ and $\bar m^\alpha$, cf.\ Eqs.\ (\ref{eq:nNP}) and (\ref{eq:mbarNP}).  Their forms are somewhat lengthy, so we present them in Appendix \ref{app:lengthy}.  An important point to note here is that these terms involve polynomial functions of $r$, powers of $\cos\theta$ and $\sin\theta$, and the coordinate velocity components $dr/d\lambda$ and $d\theta/d\lambda$.  It is also worth noting that the homogeneous solutions $R^{\rm in,up}_{lm\omega}(r)$ and their derivatives can vary significantly over the domain of an eccentric orbit's radial motion.

Because of this, the integrand function $(dt/d\lambda)I^{\infty,{\rm H}}_{lm\omega}$ can vary tremendously over its domain.  Especially for high mode number, the functions which multiply this combination make the integrand oscillate rapidly.  The function we are integrating thus can take the form of a very rapid oscillation, modulated by an envelope which, though much more slowly varying, nonetheless changes by orders of magnitude over the integration domain.

We use Clenshaw-Curtis quadrature (see, e.g., Ref.\ \cite{NumRec} for discussion) to evaluate Eq.\ (\ref{eq:Zlmkn_integral}).  This technique decomposes the integrand onto a spectral basis of Chebyshev polynomials, and evaluates the integral using analytic rules for integrating these polynomials.  It is very well suited for evaluating integrals of rapidly oscillating functions.  Because the algorithm for finding the Chebyshev expansion coefficients is essentially a Cosine Fourier Transform, this expansion can be done very effectively with Fast Fourier Transform techniques.  We use an adaptive variant of Clenshaw-Curtis quadrature, doubling the number of basis functions used in the decomposition until the fractional change in the integral falls below a specified tolerance.  We find that increasing the number of Chebyshev basis functions improves precision up to a certain point.  Beyond some maximum (typically about $2^{17} = 131072$), adding basis functions fails to improve numerical precision: at least at double precision, basis coefficients beyond this point tend to be noise dominated.  As we show in the next section, this ``noise floor'' limits the range of parameter space over which a naive application of this technique to Eq.\ (\ref{eq:Zlmkn_integral}) is useful.

\section{The ``standard'' solution, and its breakdown at large eccentricity}
\label{sec:standardcalc}

Here we examine the behavior of BHPT solutions found by implementing the procedure described in Sec.\ \ref{sec:synopsis}.  In order to focus strictly on the influence of eccentricity, we fix other variables describing the system: we put $a = 0$, $p = 8.5M$, and $x_I = 1$, and focus on modes with $l = m = 2$.  Note that for equatorial systems with $x_I = \pm 1$, only modes with $k = 0$ contribute, so we omit the $k$ index when discussing amplitudes and fluxes, labeling modes by $(2,2,n)$.

\begin{figure}[h]
\includegraphics[width=0.48\textwidth]{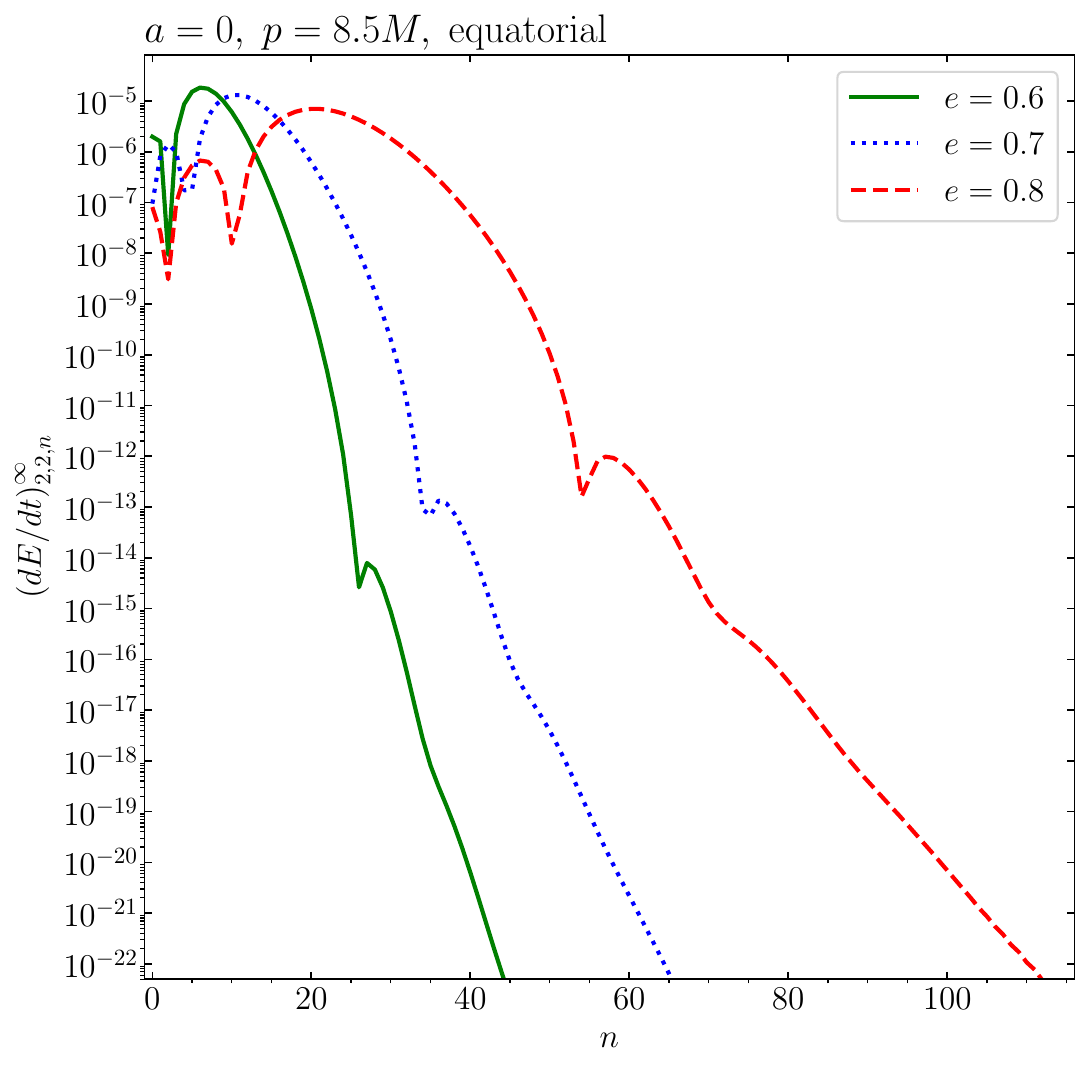}
\caption{Falloff of the energy carried by gravitational waves to infinity in the $(2,2,n)$ mode for an equatorial orbit about a Schwarzschild black hole with $p = 8.5M$ and eccentricity $e = 0.6$, $0.7$, $0.8$.  As eccentricity grows, the value of $n$ corresponding to the peak flux grows, and the slope of the high-$n$ exponential falloff becomes more shallow.}
\label{fig:falloff_norm}
\end{figure}

Figure \ref{fig:falloff_norm} shows typical examples of gravitational-wave energy flux spectra from ``well-behaved'' BHPT solutions.  We plot the contribution of the $(2,2,n)$ mode to the flux of energy carried to infinity by gravitational waves from this orbit.  To compute this, we solved for the mode amplitudes $Z^\infty_{2,2,n}$ following the procedure described in Sec.\ \ref{sec:synopsis}, and then computed the energy flux from these modes using Eq.\ (\ref{eq:dEdtinf}).  A key feature we wish to highlight is that the energy flux has several local maxima but for ``large'' $n$ decays exponentially with $n$.  The location of these features and the meaning of ``large'' varies with eccentricity: for $e = 0.6$, the peak value of $(dE/dt)^\infty_{2,2,n}$ is at $n = 6$, and the contribution has fallen by 16 orders of magnitude at $n = 40$; for $e = 0.7$, the peak is at $n = 11$, and has fallen by 16 orders of magnitude at $n = 61$; for $e = 0.8$, the peak is at $n = 20$, and has fallen by 16 orders of magnitude at $n = 108$.  Although quantitative details like the location of maxima and the slope of falloff tend to vary, very similar behavior is seen for the flux down the horizon, computed from $Z^{\rm H}_{2,2,n}$ using Eq.\ (\ref{eq:dEdthrz}), as well as for fluxes of angular momentum, and for fluxes from different orbits (varying black hole spin $a$, semi-latus rectum $p$, and inclination angle $x_I$).

\begin{figure}[h]
\includegraphics[width=0.48\textwidth]{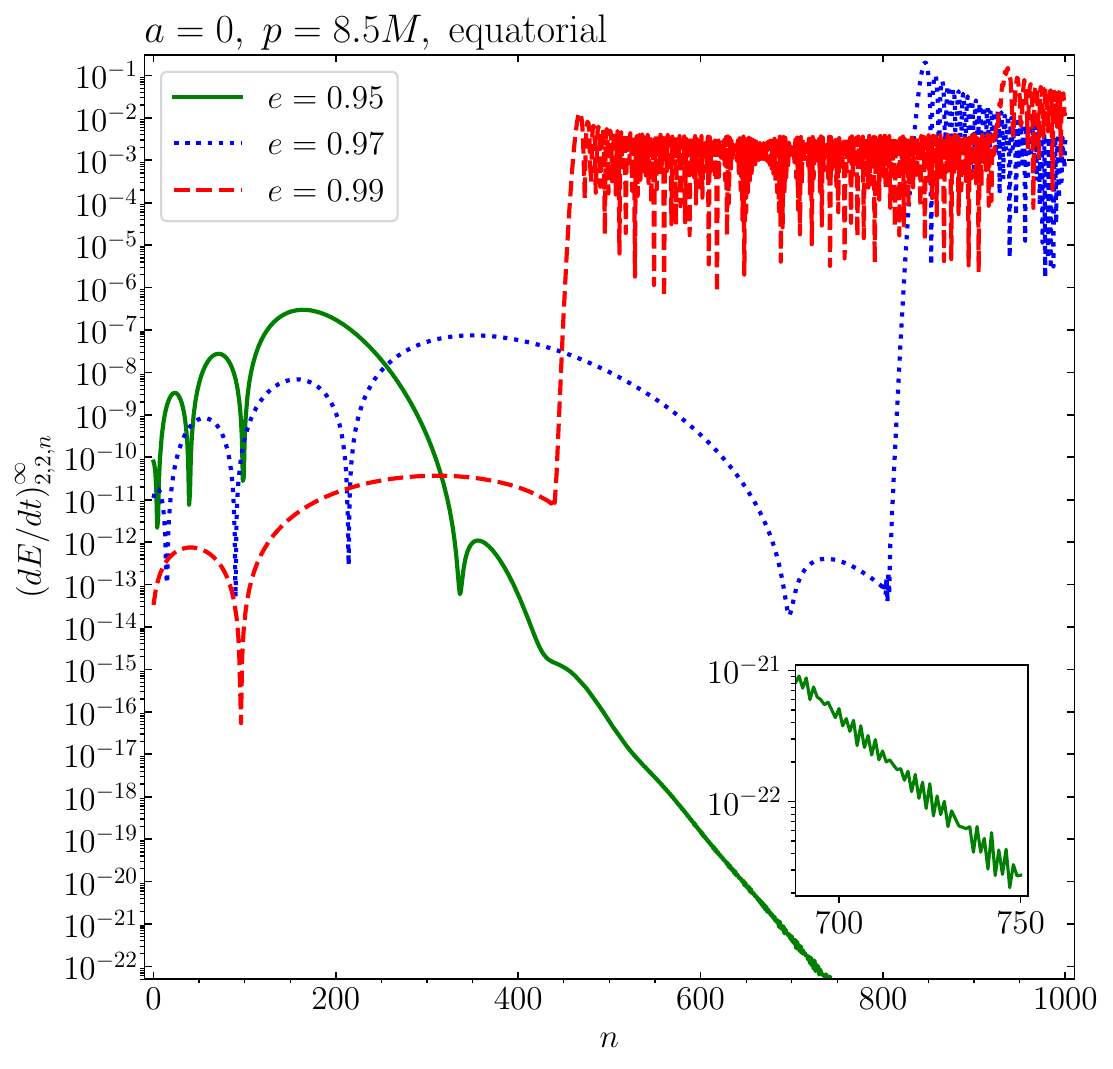}
\caption{Similar to Fig.\ \ref{fig:falloff_norm}, but now for eccentricity $e = 0.95$, $0.97$, $0.99$.  Notice that noise at $n \gtrsim 650$ is beginning to affect the spectrum for $e = 0.95$; noise dominates the calculation for $e = 0.97$ at $n \gtrsim 800$ and for $e = 0.99$ at $n \gtrsim 450$.}
\label{fig:falloff_bad}
\end{figure}

Figure \ref{fig:falloff_bad} shows examples illustrating how the most straightforward implementation of this method fails for high $e$ and high $n$.  All parameters are exactly as in Fig.\ \ref{fig:falloff_norm} except eccentricity, which we increase.  Broadly speaking, the trends described for Fig.\ \ref{fig:falloff_norm} continue here: the location of the spectral peak moves to larger $n$, and (at least for $e = 0.95$), the high-$n$ falloff becomes more shallow.  Notice that for $e = 0.95$, we find noise at the largest values of $n$ included on this plot (highlighted in the inset).

This noise is a characteristic of the high $e$, high $n$ limit of this calculation.  As $e$ increases, the value of $n$ at which the noise becomes noticeable becomes smaller: for $e = 0.97$, the noise significantly impacts our analysis at $n \gtrsim 800$; for $e = 0.99$, its effects are important at $n \gtrsim 450$.  At least in the $e = 0.95$ case, noise is small enough that its impact on the total flux calculation is small.  For $e = 0.97$ and $e = 0.99$ the noise dominates our computation, preventing us from reliably computing gravitational waves from these orbits.  Similar behaviors are seen in the flux down the horizon and for different orbits, in some cases badly affecting our analysis at smaller values of $n$.  Note that increasing the number of Chebyshev polynomials used for the Clenshaw-Curtis evaluation of (\ref{eq:Zlmkn_integral}) does not improve our results.  We appear to have hit the limit of a straightforward application of this technique, at least working with double-precision numerical methods.

\begin{figure}[h]
\includegraphics[width=0.48\textwidth]{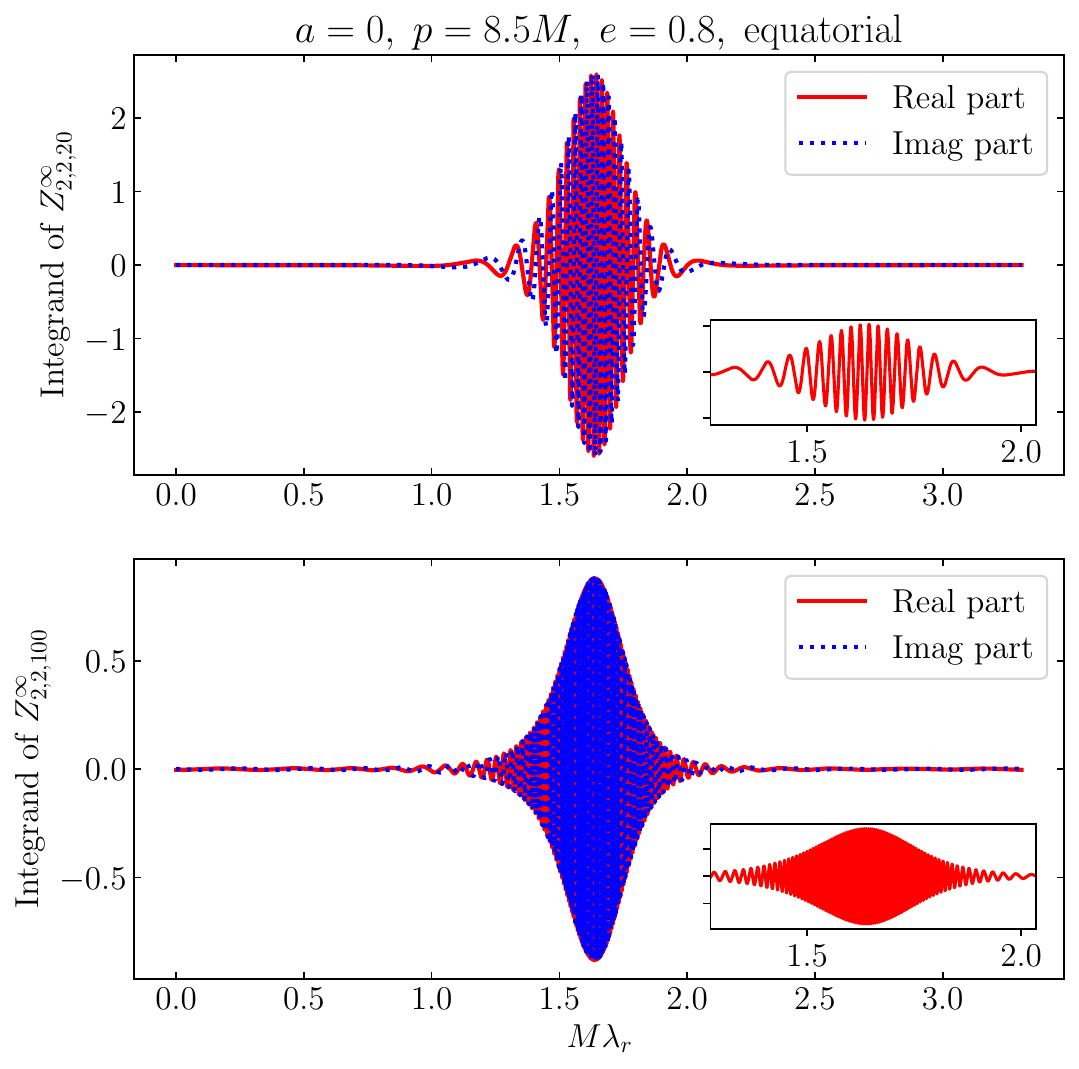}
\caption{The integrand of Eq.\ (\ref{eq:Zlmkn_integral}) for $l = 2$, $m = 2$, and for an equatorial orbit with $a = 0$, $p = 8.5M$, $e = 0.8$, plotted from $0 \le \lambda_r \le \Lambda_r$, where $\Lambda_r = 3.3087/M$.  Upper panel is for $n = 20$, lower is for $n = 100$.  Notice that the integrand takes the form of an envelope which modulates a highly oscillatory function.  This envelope is practically flat over the range $0 \le M\lambda_r \lesssim 1.2$, rapidly grows and rapidly falls over the range $1.2 \lesssim M\lambda_r \lesssim 2.1$ (peaking near $\lambda_r = \Lambda_r/2$), and again is practically flat over the range $2.1 \lesssim M\lambda_r \le M\Lambda_r$.  At its peak, the envelope is larger than its minimum value by a factor of about $1390$ for $n = 20$, and by a factor of about $340$ for $n = 100$.  The integrand varies quite slowly over the nearly flat region, but rapidly oscillates over the range in which the envelope grows and falls.  The inset zooms into a region of width $\Delta(M\lambda_r) = 0.76$ centered on the peak.  The numerical challenge is to accurately sum the rapidly varying contributions which accumulate under this more slowly varying envelope and nearly cancel out.}
\label{fig:igrnd_good}
\end{figure}

To understand why our numerical results appear to be reliable for the cases shown in Fig.\ \ref{fig:falloff_norm} but not for those shown in Fig.\ \ref{fig:falloff_bad}, we now examine the integrand of Eq.\ (\ref{eq:Zlmkn_integral}) for several representative modes.  Figure \ref{fig:igrnd_good} shows the integrand for two of the points which contribute to the $e = 0.8$ spectrum in Fig.\ \ref{fig:falloff_norm}: $n = 20$ (upper panel), and $n = 100$ (lower).  For this orbit, $\Lambda_r = 3.3087/M$; we show the integrand over the range $0 \le \lambda_r \le \Lambda_r$.  In both cases, we see that the integrand is quite flat over the range $0 \le M\lambda_r \lesssim 1.2$ and $2.1 \lesssim M\lambda_r \le M\Lambda_r$, but rises and falls rapidly in the intermediate region $1.2 \lesssim M\lambda_r \lesssim 2.1$.  During this rapid rise and fall, the envelope of the integrand changes by several orders of magnitude from its value in the nearly flat region (increasing by a factor of about $1390$ for $n = 20$, and by a factor of about $340$ for $n = 100$).  The integrand oscillates many times, completing approximately $n + m$ cycles over its complete range, with the oscillations almost entirely confined to the intermediate region of rapid rise and fall.  Notice that the oscillations are not of constant period: the period is much longer near the edge of the rapid rise and fall than near the central peak.

For the  cases shown in Fig.\ \ref{fig:igrnd_good}, we find
\begin{align}
    Z^\infty_{2,2,20} &= \left(-1.4845 + 0.46998i\right)\times 10^{-3}\;,
    \nonumber\\
    Z^\infty_{2,2,100} &= -\left(1.9291 + 0.87197i\right)\times 10^{-10}\;.
\end{align}
In both these cases the value of the integral is much smaller than the integrand's amplitude.  Because of the many rapid oscillations under the more slowly varying envelope, this function nearly integrates to zero over the range $0$ to $\Lambda_r$.

\begin{figure}[h]
\includegraphics[width=0.457\textwidth]{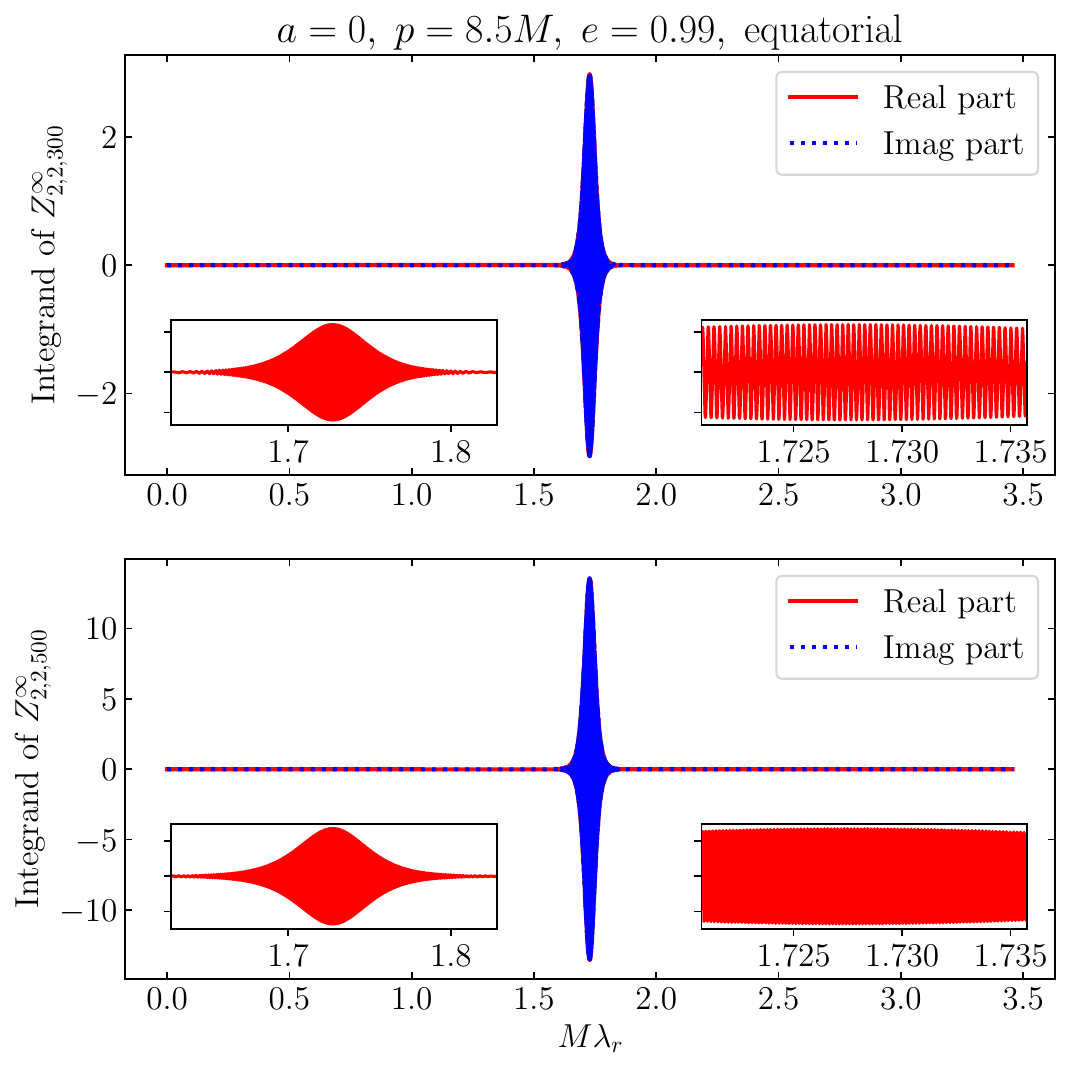}
\caption{Similar to Fig.\ \ref{fig:igrnd_good}, but now for $e = 0.99$.  For this case, $\Lambda_r = 3.4565/M$.  Upper panel is for $n = 300$, lower is for $n = 500$.  Though these integrands are qualitatively similar to the cases examined in Fig.\ \ref{fig:igrnd_good}, several behaviors are significantly more extreme: the range over which the integrand rapidly oscillates, $1.6 \lesssim M\lambda_r \lesssim 1.8$, is much narrower than for the cases shown in Fig.\ \ref{fig:igrnd_good}; the integrand undergoes many more oscillations over this narrow range; the ratio of envelope peak to minimum is about $9\times 10^7$ for $n = 300$, and is about $5 \times 10^7$ for $n = 500$.  The left-hand inset panels zoom onto a range $\Delta(M\lambda_r) = 0.2$ centered on the peak; the right inset panels zoom onto a likewise centered range of $\Delta(M\lambda_r) = 0.015$.  The oscillations are so dense that they can barely be discerned at the most extreme zoom shown in the right-hand insets.  The numerical challenge of accurately summing these rapidly varying contributions is quite extreme in these cases.}
\label{fig:igrnd_bad}
\end{figure}

The Clenshaw-Curtis algorithm that we use, as described in Sec.\ \ref{sec:source}, is well suited for problems of this sort, allowing us to accurately determine integrals for functions that rapidly oscillate under a more slowly varying, high-amplitude envelope.  However, a straightforward application of this method fails when the oscillations become too dense and rapid.  Figure \ref{fig:igrnd_bad} shows the integrand for two of the points which contribute to the $e = 0.99$ spectrum in Fig.\ \ref{fig:falloff_bad}: $n = 300$ (upper panel) and $n = 500$ (lower panel).  Note that the spectrum appears to be well-behaved at $n = 300$, but is in the noise-dominated regime at $n = 500$.  The integrands in these cases share many qualitative features of the integrands shown in Fig.\ \ref{fig:igrnd_good}, but here taken to a much greater extreme: the rapidly rising and falling peak is quite a bit narrower; the number of oscillations is much larger; and the contrast between the integrand's amplitude at peak versus near its minimum is far larger (increasing by a factor of about $9 \times 10^7$ for $n = 300$, and $5 \times 10^7$ for $n = 500$).  Despite the much greater challenge of accurately resolving these rapid oscillations, the Clenshaw-Curtis algorithm succeeds for $n = 300$.  The greater number of oscillations for $n = 500$ appears to overwhelm the capabilities of the double-precision code we use.

\section{Precisely evaluating the source integral for large eccentricity}
\label{sec:high_e_calc}

Our goal now is to develop a method of reliably evaluating Eq.\ (\ref{eq:Zlmkn_integral}) as $e \to 1$ for very large values of $n$.  In particular, we would like to insure that we can evaluate these quantities to sufficiently large values of $n$, such that quantities like the energy flux $(dE/dt)^{\infty,{\rm H}}_{lmn}$ exhibit the exponential falloff that we saw for modest eccentricity.  We describe two techniques for ameliorating the challenge associated with the high $e$, large $n$ limit.  In the end, we find that the simplest of these techniques appears to be adequate for current analyses.  The more sophisticated analysis we describe may be useful for very extreme situations, though it incurs significantly greater computational cost.

\subsection{Subdividing the interval}
\label{sec:be_cool_or_be_cast_out}

Figures \ref{fig:igrnd_good} and \ref{fig:igrnd_bad} illustrate the challenging numerical properties of the integral (\ref{eq:Zlmkn_integral}).  Although the spectral Clenshaw-Curtis method is well-suited for integrating rapidly oscillating functions of the sort which appear here, for sufficiently large $n$, the integrand's challenges overwhelm a straightforward application of this method.

\begin{figure*}[ht]
\includegraphics[width=0.30\textwidth]{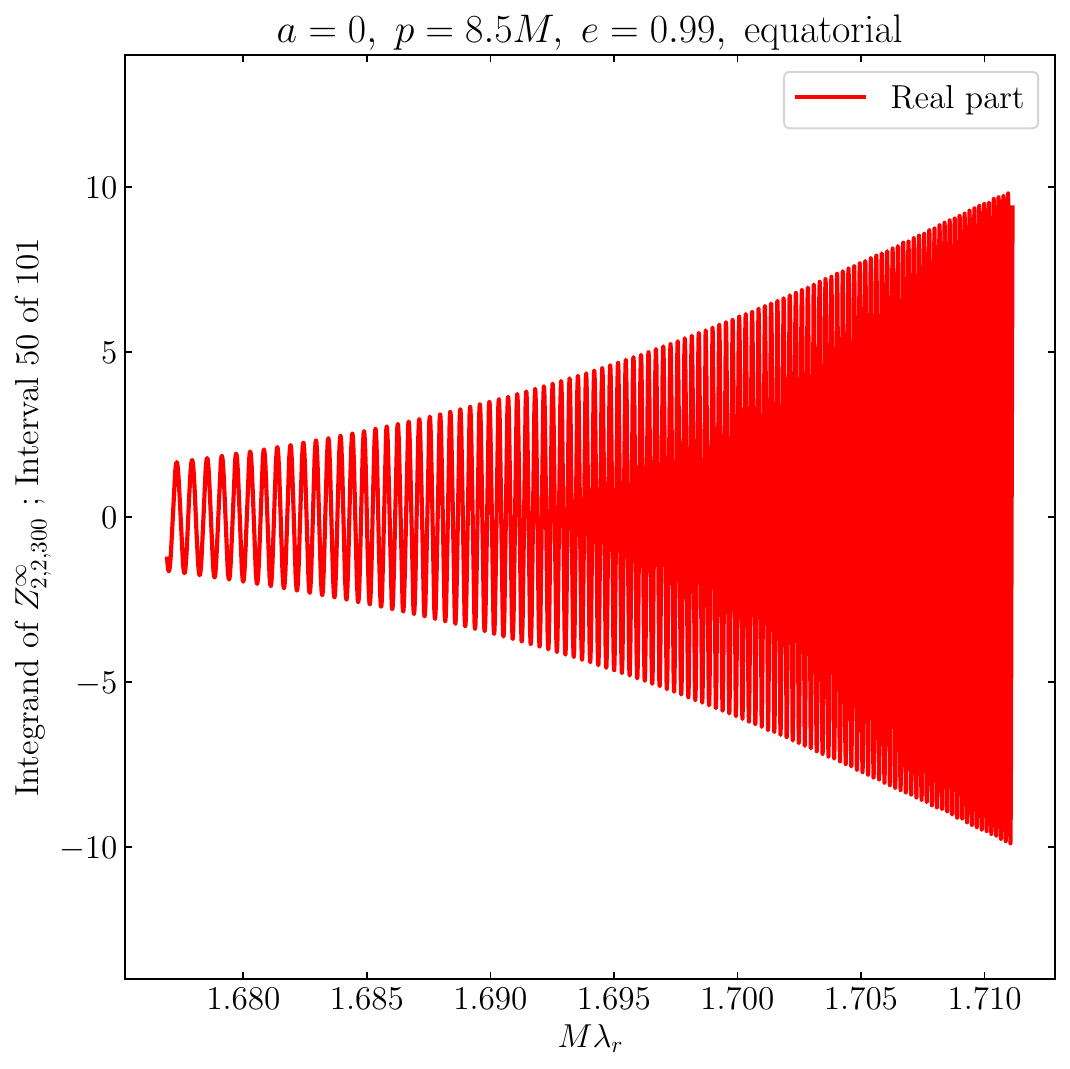}
\includegraphics[width=0.30\textwidth]{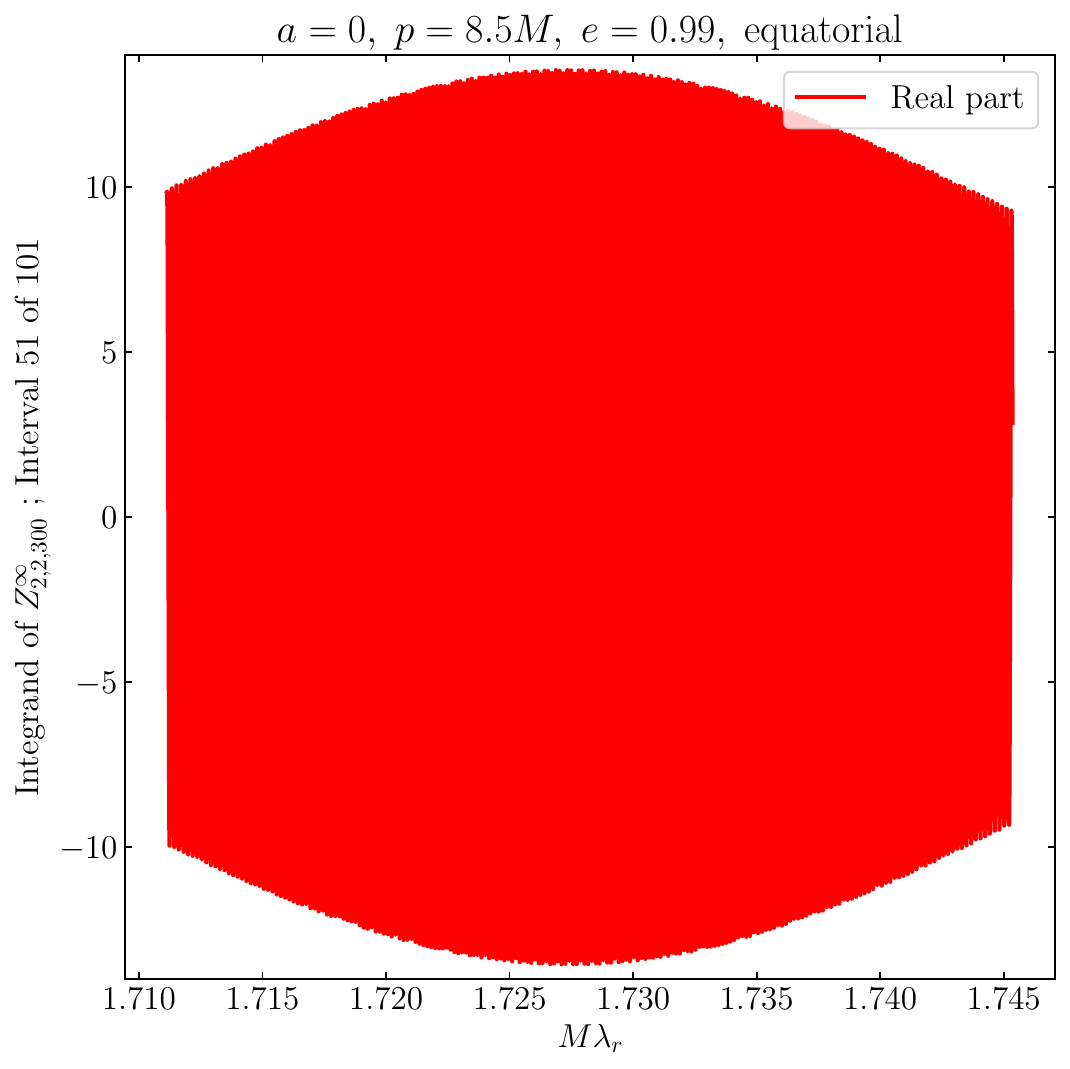}
\includegraphics[width=0.30\textwidth]{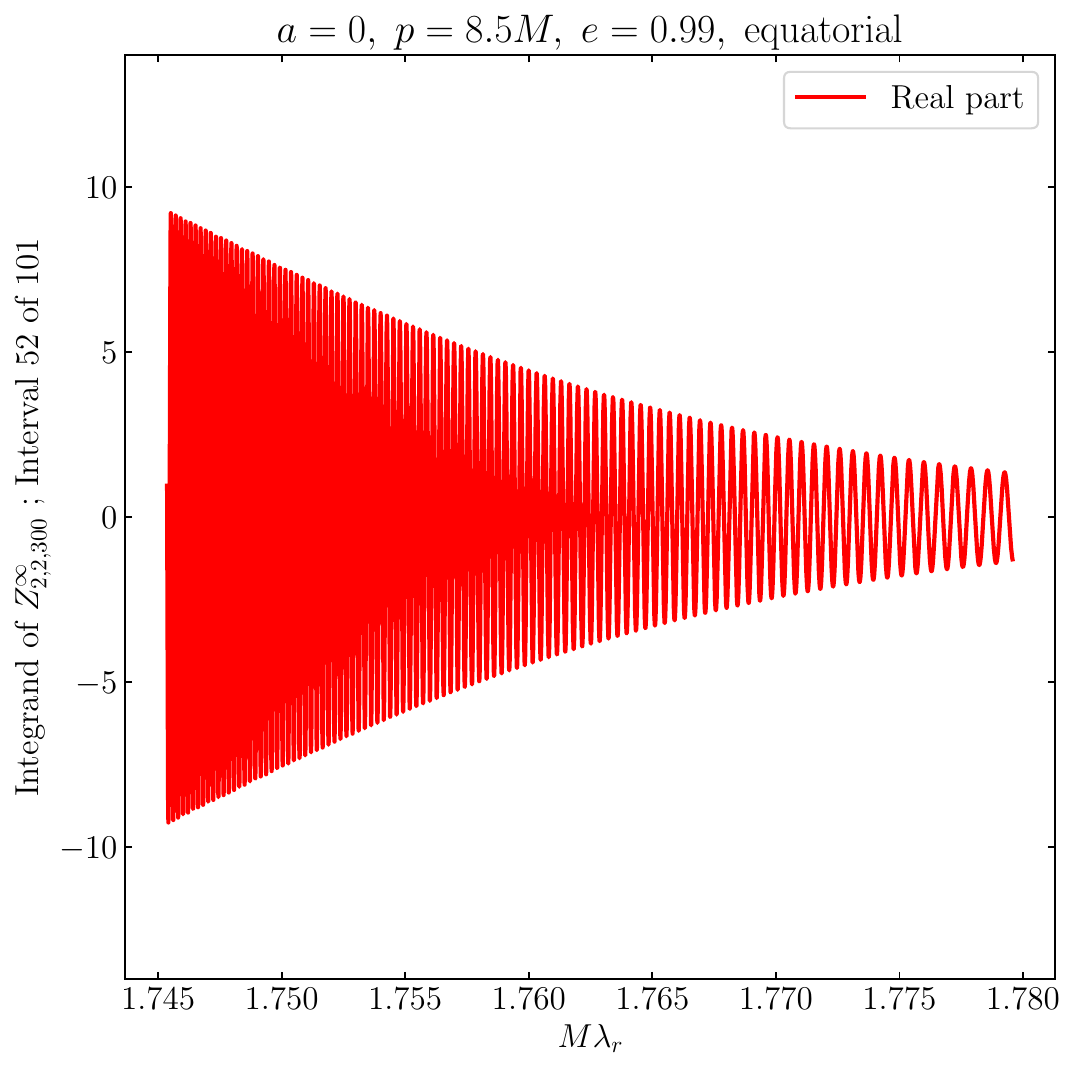}
\caption{Three intervals near the center of the integrand shown in Fig.\ \ref{fig:igrnd_bad}, after dividing into $N_{\rm int} = 101$ total intervals.  Left-hand panel shows the $50$th interval, covering the region $(49/101)\Lambda_r \le \lambda_r < (50/101)\Lambda_r$; center panel shows the $51$st interval, covering $(50/101)\Lambda_r \le \lambda_r < (51/101)\Lambda_r$; right-hand shows the $52$nd interval, covering $(51/101)\Lambda_r \le \lambda_r < (52/101)\Lambda_r$.  The integrand oscillates approximately 105 times in the left- and right-hand panels; it oscillates about 205 times in the center panel.  Reducing the total number of oscillations in each integration region, as well as the contrast between the minimum and maximum value of the integrand's envelope in each region, significantly reduces the numerical challenge of accurately evaluating $Z^{\infty,{\rm H}}_{lmn}$.}
\label{fig:subdivide}
\end{figure*}

The most straightforward way to address this challenge is to reduce the number of oscillations which appear in the integration region.  We break the integral over $0 \le \lambda_r \le \Lambda_r$ into $N_{\rm int}$ intervals, integrating over
\begin{align}
0 \le \lambda_r < \Lambda_r/N_{\rm int}\qquad&\mbox{interval 1}\;,
\nonumber\\
\Lambda_r/N_{\rm int} \le \lambda_r < 2\Lambda_r/N_{\rm int}\qquad&\mbox{interval 2}\;,
\nonumber\\
\ldots
\nonumber\\
(N_{\rm int} - 1)\Lambda_r/N_{\rm int} \le \lambda_r \le \Lambda_r\qquad&\mbox{interval $N_{\rm int}$}\;.
\end{align}
We then sum the contributions from each interval to compute Eq.\ (\ref{eq:Zlmkn_integral}).  Figure \ref{fig:subdivide} shows three subdivisions of the integrand for the $n = 500$ case shown in Fig.\ \ref{fig:igrnd_bad}, using $N_{\rm int} = 101$, displaying intervals $50$, $51$, and $52$. 

\begin{figure}[h]
\includegraphics[width=0.48\textwidth]{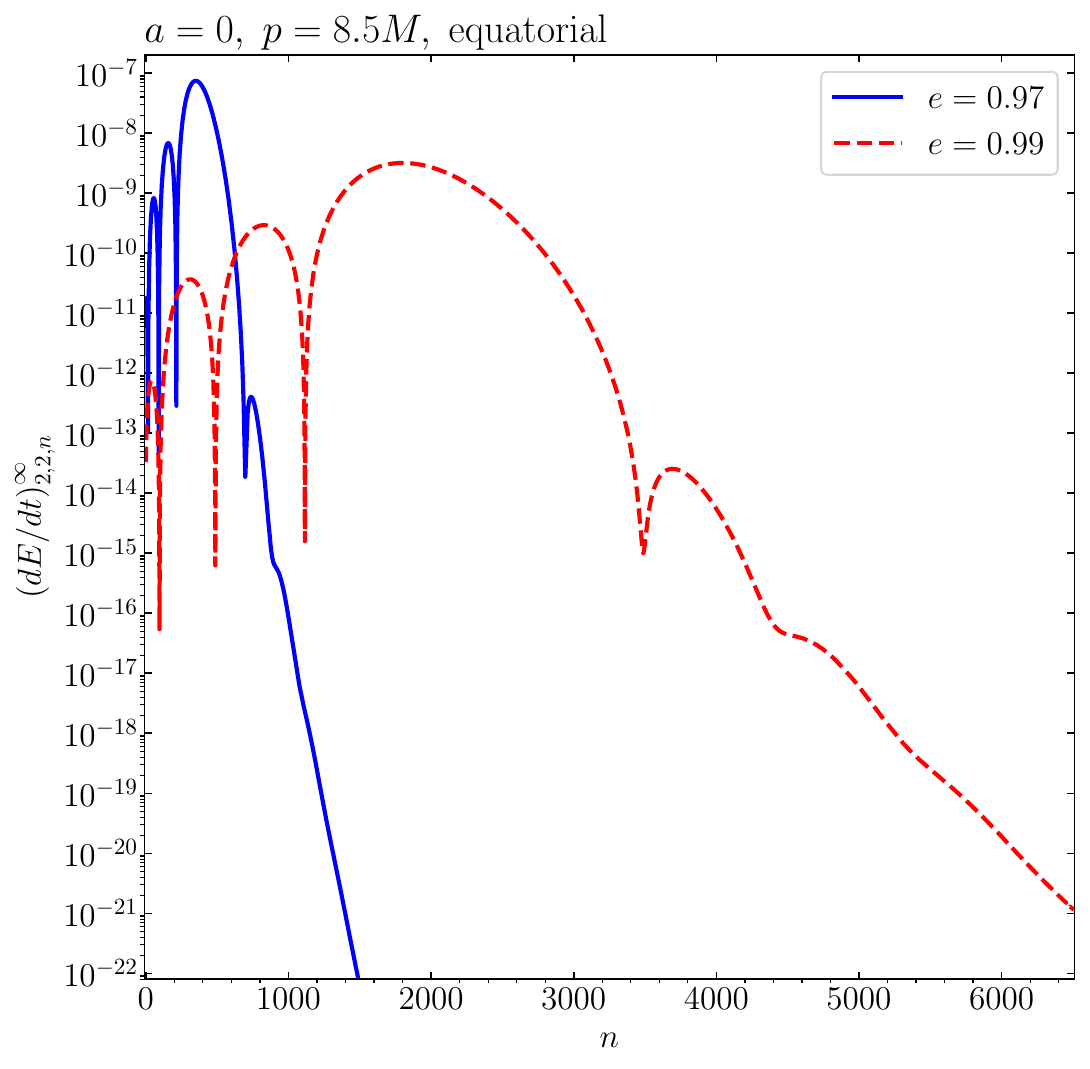}
\caption{The falloff of energy carried by gravitational waves to infinity in the $(2,2,n)$ mode for equatorial orbits about a Schwarzschild black hole with $p = 8.5M$.  We show results for $e = 0.97$ and $e = 0.99$, the two cases which are dominated by noise at large $n$ in Fig.\ \ref{fig:falloff_bad}.  We now evaluate the source integral (\ref{eq:Zlmkn_integral}) using the subdivided interval method with $N_{\rm int} = 101$, as described in Sec.\ \ref{sec:be_cool_or_be_cast_out}.  Doing so, we find that the high $n$ noise is completely eliminated, at least over the range required for the spectrum to show exponential fall off for these cases.  Although not shown here, the energy flux into the horizon is also well behaved over this entire range of $n$.}
\label{fig:falloff_fixed}
\end{figure}

Figure \ref{fig:falloff_fixed} shows how this simple adjustment vastly improves the evaluation of Eq.\ (\ref{eq:Zlmkn_integral}).  We show here the two cases from Fig.\ \ref{fig:falloff_bad} that are dominated by noise, but now evaluating the source integral using the subdivided interval method described above, with $N_{\rm int} = 101$.  The $l = m = 2$ spectrum for $(a, p, e) = (0, 8.5M, 0.97)$ shows local peaks at $n = 4$, $n \simeq 54$, $155$, $350$ (which is also a global peak), and $740$, as well as a small bump at $n \simeq 940$.  The spectrum then behaves like an exponential falloff through $n \gtrsim 1500$, at which point it has fallen by about 15 orders of magnitude from its peak.  The $l = m = 2$ spectrum for $(a, p, e) = (0, 8.5M, 0.99)$ shows local peaks at $n = 41$, $n \simeq 309$, $820$, $1830$ (the global peak), and $3700$, with a small bump at $n \simeq 4660$.  The spectrum then falls off exponentially through $n \gtrsim 6500$, at which point it has fallen by about 12 orders of magnitude from its peak.

\begin{figure}[h]
\includegraphics[width=0.48\textwidth]{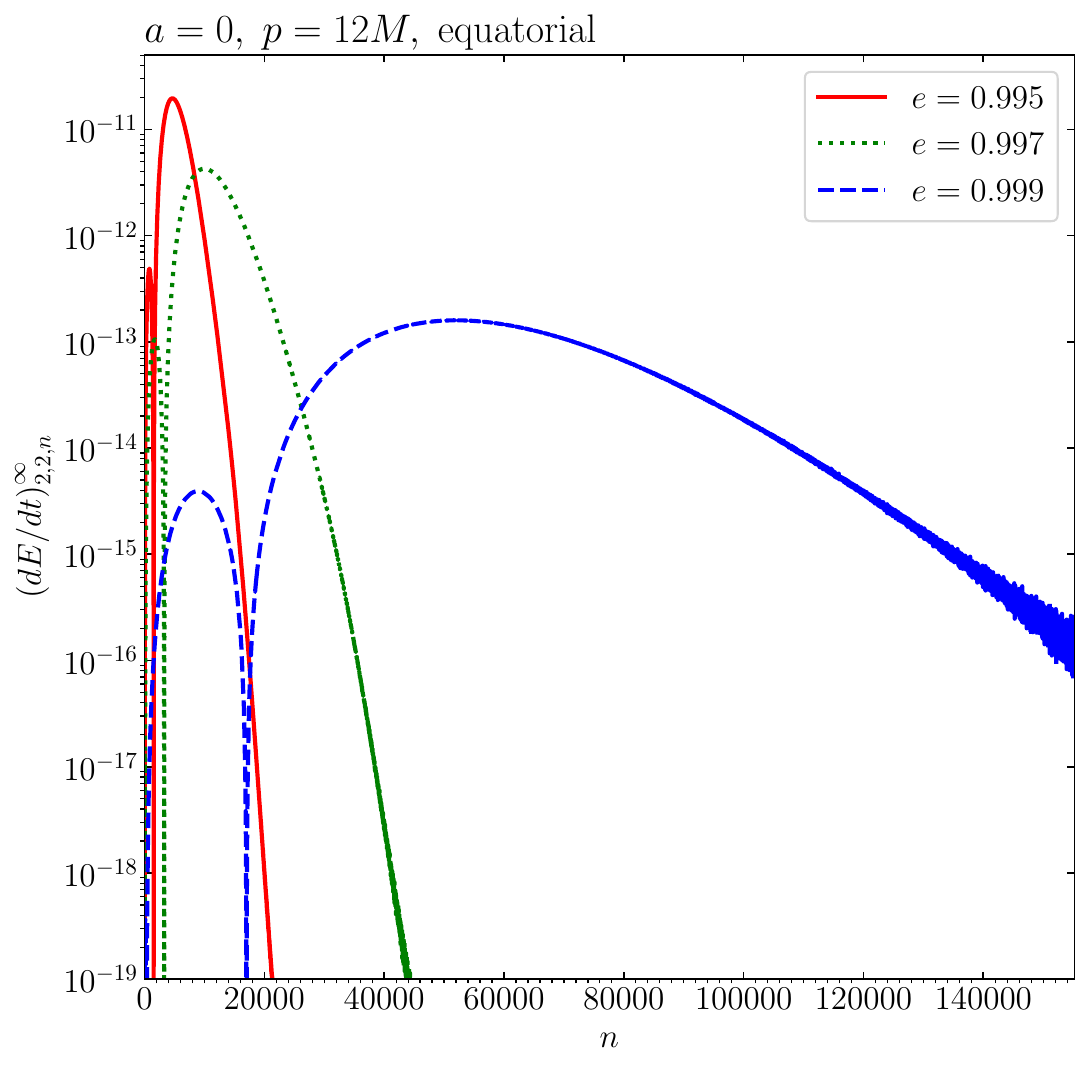}
\caption{The falloff of energy carried by gravitational waves to infinity in the $(2,2,n)$ mode for equatorial orbits about a Schwarzschild black hole with $p = 12M$.  We show results for $e = 0.995$, $e = 0.997$, and $e = 0.999$.  Very similar behavior is seen for the energy flux into the horizon for these orbits.  We see that subdividing the integral (using $N_{\rm int} = 101$ intervals for $e = 0.995$, $N_{\rm int} = 501$ for $e = 0.997$, and $N_{\rm int} = 1201$ for $e = 0.999$) allows us to precisely compute these spectra, at least for $n \lesssim 120000$.  We have been unable to eliminate the high-$n$ noise which begins to significantly affect the spectrum for $e = 0.999$ at $n \simeq 120000$, and speculate that this may represent the limit of these frequency-domain techniques, at least using double-precision methods.}
\label{fig:falloff_highecc}
\end{figure}

Figure \ref{fig:falloff_highecc} looks at the $(2,2,n)$ spectrum for $a = 0$, $p = 12M$, and several values of $e$.  We see that these methods can be used to compute spectra quite precisely at least to $e = 0.997$.  In this case, subdividing the integral into $N_{\rm int} = 501$ intervals, we find a high-quality spectrum out to $n \simeq 40000$, although small-scale noise artifacts begin to impact the spectrum for the highest radial harmonics.  Increasing to $N_{\rm int} = 1201$, we are able to compute much of the spectrum for $e = 0.999$.  It transitions to an exponential falloff when $n \gtrsim 70000$, though we find that noise makes a substantial contribution to the spectrum for $n \gtrsim 120000$.  We have been unable to reduce the noise in this portion of the spectrum for $e = 0.999$.

Subdividing the interval as described here appears to be adequate for accurately evaluating Eq.\ (\ref{eq:Zlmkn_integral}) in all the cases we have studied so far, out to $e \simeq 0.997$.  It may be possible to improve precision somewhat by implementing a more sophisticated analysis of the source integral which uses integration by parts to separate contributions to the integrand (\ref{eq:source_integrand}) which vary on different scales.  We next describe this idea.

\subsection{Integrating by parts}
\label{sec:byparts}

As already emphasized, the fundamental numerical challenge when attempting to evaluate Eq.\ (\ref{eq:Zlmkn_integral}) as $e \to 1$ and for $n$ large is that the integrand consists of a rapidly oscillating function, modulated by an envelope which varies more slowly, though over several orders of magnitude.  Such a situation is an ideal set up for integration by parts, a technique that has been applied to the source integral in similar problems related to studies of the self force on unbound scattering orbits \cite{Whittall:2023xjp}.  Focusing on the equatorial case (for which only $k = 0$ contributes, so we drop the $k$ index), we begin by rewriting the source integral (\ref{eq:Zlmkn_integral}) as
\begin{equation}
Z^{\infty,{\rm H}}_{lmn} = \frac{1}{\Lambda_r}\int_0^{\Lambda_r} \mathcal{I}^{\infty,{\rm H}}_{lmn}(\lambda_r)d\lambda_r\;.
\label{eq:Zlmn_rewrite}
\end{equation}
The integrand function $\mathcal{I}^{\infty,{\rm H}}_{lmn}(\lambda_r)$ can be read out of Eqs.\ (\ref{eq:Zlmkn_integral}) and (\ref{eq:source_integrand}).  Let us define the envelope function
\begin{equation}
    {{_0}\mathcal{E}}^{\infty,{\rm H}}_{lmn}(\lambda_r) = \sqrt{\left(\mbox{Re}(\mathcal{I}^{\infty,{\rm H}}_{lmn})\right)^2 + \left(\mbox{Im}(\mathcal{I}^{\infty,{\rm H}}_{lmn})\right)^2}\;,
    \label{eq:envelope}
\end{equation}
and the rapidly varying complex sinusoid
\begin{align}
    {{_0}\mathcal{S}}^{\infty,{\rm H}}_{lmn}(\lambda_r) &= e^{i\phi^{\infty,{\rm H}}_{lmn}(\lambda_r)}\;,\quad\mbox{where}
    \nonumber\\
    \tan\phi^{\infty,{\rm H}}_{lmn}(\lambda_r) &= \mbox{Im}(\mathcal{I}^{\infty,{\rm H}}_{lmn})/\mbox{Re}(\mathcal{I}^{\infty,{\rm H}}_{lmn})\;.
    \label{eq:complexsinusoid}
\end{align}
Then,
\begin{align}
    Z^{\infty,{\rm H}}_{lmn} &= \frac{1}{\Lambda_r}\int_0^{\Lambda_r} {{_0}\mathcal{E}}^{\infty,{\rm H}}_{lmn}(\lambda_r) {{_0}\mathcal{S}}^{\infty,{\rm H}}_{lmn}(\lambda_r)d\lambda_r
    \label{eq:Zlmn_rewrite2}\\
    &= \frac{1}{\Lambda_r}\Biggl[{{_0}\mathcal{E}}^{\infty,{\rm H}}_{lmn}(\Lambda_r){_{-1}\mathcal{S}}^{\infty,{\rm H}}_{lmn}(\Lambda_r)
    \nonumber\\
    &\qquad - {{_0}\mathcal{E}}^{\infty,{\rm H}}_{lmn}(0) {_{-1}\mathcal{S}}^{\infty,{\rm H}}_{lmn}(0)
    \nonumber\\
    &\qquad - \int_0^{\Lambda_r} {{_1}\mathcal{E}}^{\infty,{\rm H}}_{lmn}(\lambda_r) {_{-1}\mathcal{S}}^{\infty,{\rm H}}_{lmn}(\lambda_r) d\lambda_r \Biggr]\;.
    \label{eq:Zlmn_byparts}
\end{align}
Equation (\ref{eq:Zlmn_rewrite2}) is identical to (\ref{eq:Zlmn_rewrite}), but the integral is written using the envelope ${{_0}\mathcal{E}}^{\infty,{\rm H}}_{lmn}(\lambda_r)$ and the complex sinusoid ${{_0}\mathcal{S}}^{\infty,{\rm H}}_{lmn}(\lambda)$.  We then integrate by parts, introducing the first derivative of the envelope and the first anti-derivative of the sinusoid:
\begin{align}
    {_{1}\mathcal{E}}^{\infty,{\rm H}}_{lmn}(\lambda_r) &= \frac{d{_{0}\mathcal{E}}^{\infty,{\rm H}}_{lmn}}{d\lambda_r}\;,
    \\
    {_{-1}\mathcal{S}}^{\infty,{\rm H}}_{lmn}(\lambda_r) &= \int_0^{\lambda_r} {_{0}\mathcal{S}}^{\infty,{\rm H}}_{lmn}(\lambda_r')d\lambda_r'\;.
\end{align}
Because the algorithm we use is set up to expand the integrand in a Chebyshev basis in order to do the Clenshaw-Curtis integration, it is a simple tweak to instead introduce expansions of ${_{0}\mathcal{E}}^{\infty,{\rm H}}_{lmn}$ and ${_{0}\mathcal{S}}^{\infty,{\rm H}}_{lmn}$.  Once the Chebyshev expansion of these functions is in hand, it is likewise simple to compute their derivatives and anti-derivatives.

Although in principle integrating by parts seems likely to be a useful way to approach the source integral, we have not found any operational advantage to implementing this method.  Indeed, we have actually found that the code which computes the source integrals is significantly slower than the code which implements the simpler ``subdivide the integrand'' technique, no doubt because of the many additional computations and evaluations needed to construct the envelope function and its derivative, the complex sinusoid and its anti-derivative, and then to evaluate the by-parts integral.  For example, for the orbit with $a = 0$, $p = 8.5M$, $e = 0.99$ that we have studied above, using both methods we find
\begin{equation}
    Z^\infty_{2,2,1000} = (4.03044697 - 1.34743323i) \times 10^{-6}\;.
\end{equation}
However, our code implementing integration by parts runs about 30 times slower than subdividing the integration interval (150 CPU seconds for the integration-by-parts code on an Apple M4 architecture, versus 4.7 seconds for the subdivided interval on the same hardware).  Perhaps this could be improved, and perhaps this method would allow us to reduce the high $n$ noise which eventually dominates the spectrum, as shown in Fig.\ \ref{fig:falloff_highecc}.  We leave this to future exploration.

\begin{figure*}[ht]
\includegraphics[width=0.317\textwidth]{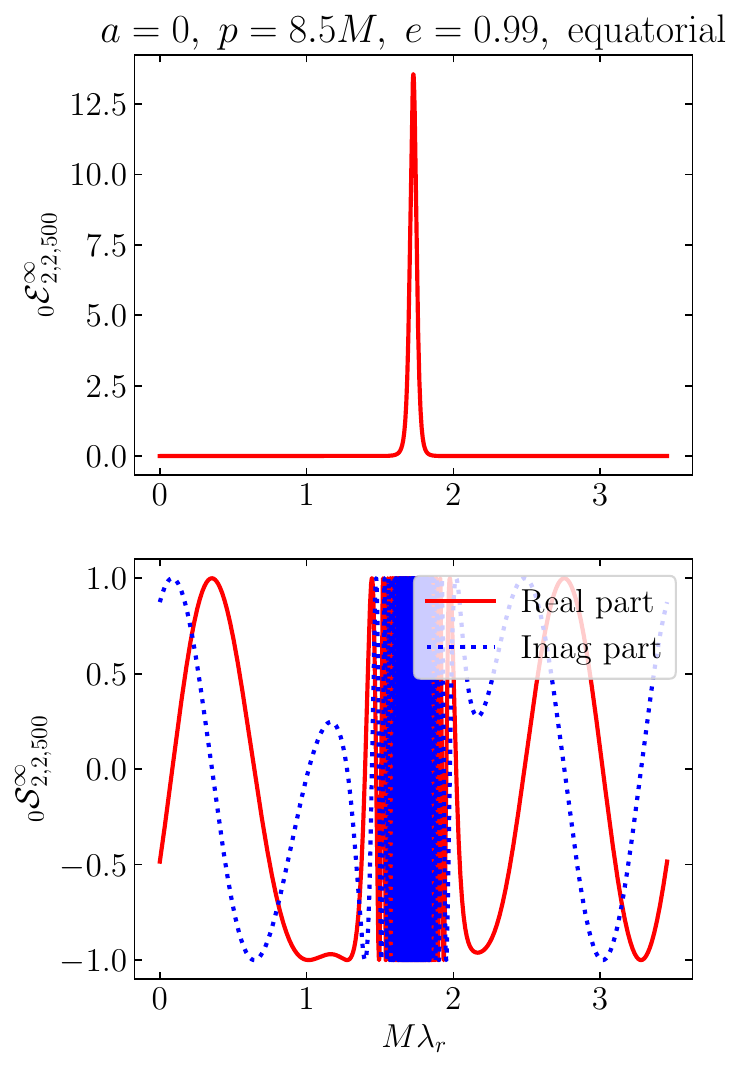}
\includegraphics[width=0.32\textwidth]{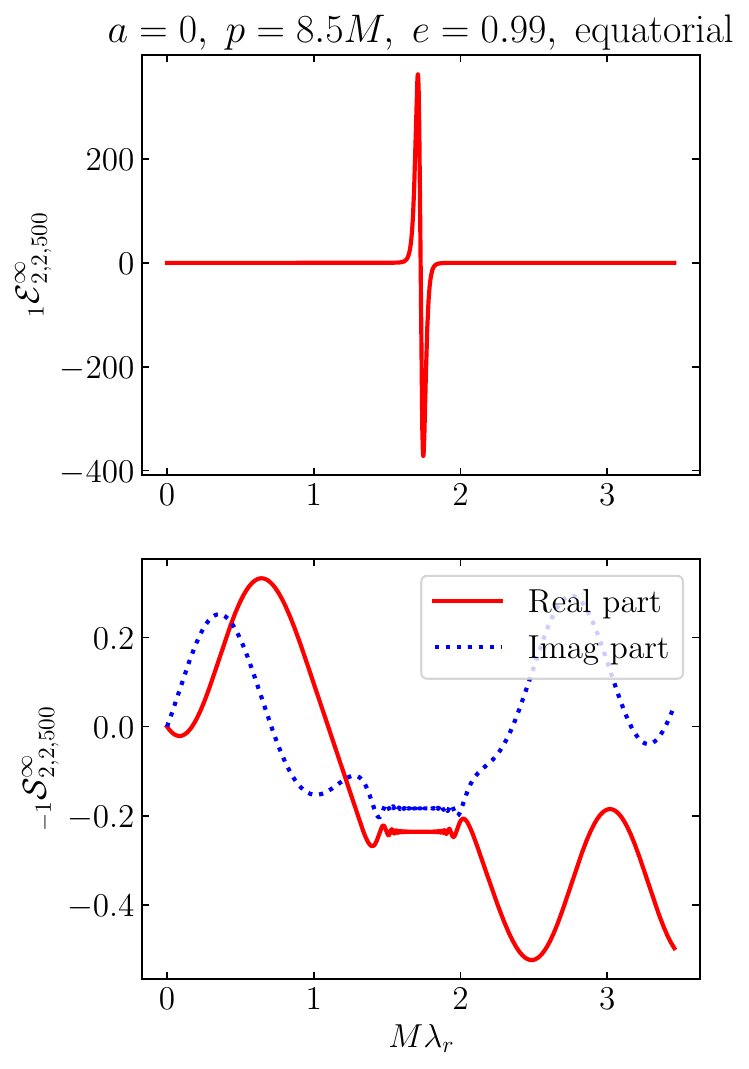}
\includegraphics[width=0.313\textwidth]{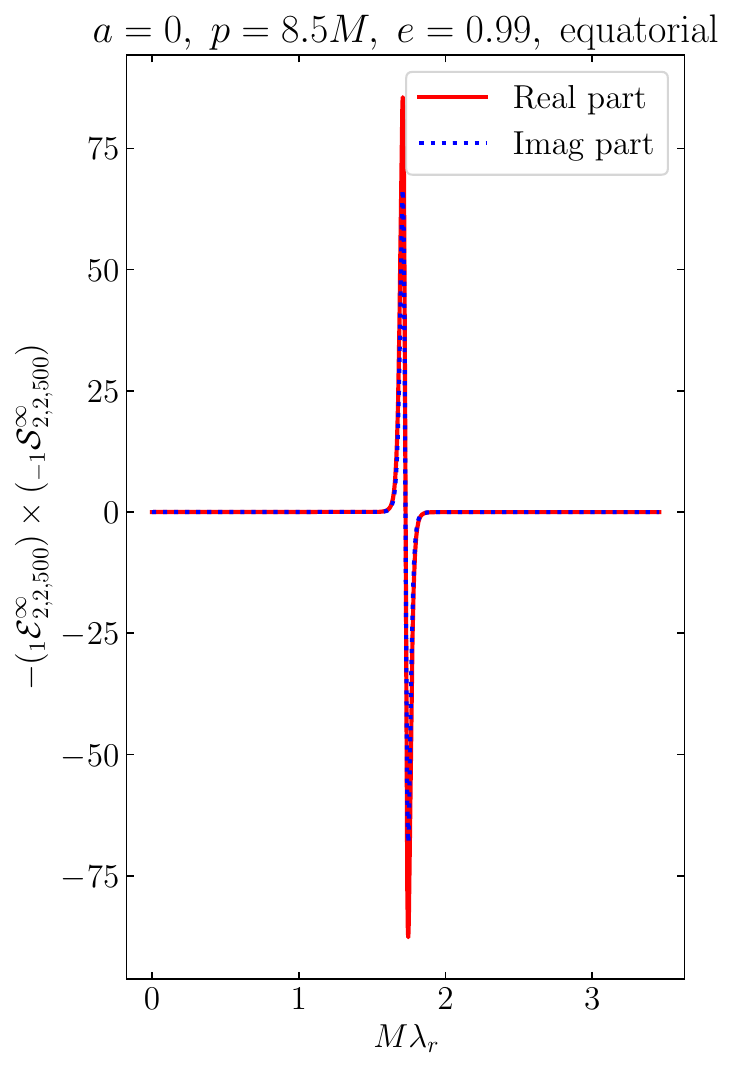}
\caption{Further analysis of the integrand for the amplitude $Z^\infty_{2,2,500}$ for the equatorial orbit with $a = 0$, $p = 8.5M$, $e = 0.99$.  Left panels: the envelope ${_{0}\mathcal{E}}^{\infty}_{2,2,500}$ (top) and the complex sinusoid ${_{0}\mathcal{S}}^{\infty}_{2,2,500}$ (bottom) as defined in Eqs.\ (\ref{eq:envelope}) and (\ref{eq:complexsinusoid}) for the integrand shown in Figs.\ \ref{fig:igrnd_bad} and \ref{fig:subdivide}.  Notice how rapidly ${_{0}\mathcal{S}}^{\infty}_{2,2,500}$ oscillates in the central region of this plot, and the sharpness of the peak of the envelope function in this region.  Center panels: the envelope derivative ${_{1}\mathcal{E}}^{\infty}_{2,2,500}$ and complex sinusoid anti-derivative ${_{-1}\mathcal{S}}^{\infty}_{2,2,500}$ which are used when evaluating this integral using integration by parts.  The anti-derivative of the sinusoid is particularly simplified here, since the integral hardly accumulates over the domain of the rapidly oscillating region.  Right panel: the product function $-({_{1}\mathcal{E}}^{\infty}_{2,2,500}) \times ({_{-1}\mathcal{S}}^{\infty}_{2,2,500})$ used when computing the amplitude $Z^\infty_{2,2,500}$ using integration by parts.  This function is far simpler to integrate than the original integrand.}
\label{fig:igrnd_byparts}
\end{figure*}

\section{Some high eccentricity gravitational wave flux results}
\label{sec:results}

With a framework in hand for precisely computing gravitational waves in the limit $e \to 1$, we briefly examine several interesting results that can be found with it.  We look at the total gravitational wave energy flux, including radiation both to infinity and down the horizon, and including enough modes to satisfy a convergence criterion (described in more detail below), from a set of Schwarzschild and Kerr ($a = 0.9M$) eccentric equatorial orbits. We very briefly compare the total fluxes from very eccentric orbits to predictions from weak-field formulas, and we examine the convergence of the total energy and angular momentum radiated as a function of $e$ in the limit $e \to 1$.  We emphasize that this discussion is not intended to be comprehensive; indeed, several of the results we show merit much more detailed study.

\subsection{Converged fluxes for a variety of orbits}
\label{sec:convergedfluxes}

The total energy flux from a black hole orbit, is found by summing the contributions from each mode:
\begin{align}
    \left(\frac{dE}{dt}\right)^{\rm tot} &= \sum_{n = -\infty}^\infty \dot E_n\;,
    \label{eq:dEdttot1}\\
    \dot E_n &= \sum_{k = -\infty}^\infty \dot E_k\;,
    \label{eq:dEdttot2}\\
    \dot E_k &= \sum_{m = -\infty}^\infty \dot E_m\;,
    \label{eq:dEdttot3}\\
    \dot E_m &= \sum_{l = l_{\rm min}}^\infty \left(\dot E^\infty_{lmkn} + \dot E^{\rm H}_{lmkn}\right)\;.
    \label{eq:dEdttot4}
\end{align}
The terms $\dot E^{\infty,{\rm H}}_{lmkn}$ are the summands appearing in Eqs.\ (\ref{eq:dEdtinf}) and (\ref{eq:dEdthrz}), respectively.  The value $l_{\rm min} = {\rm max}(2, |m|)$.  Analogous formulas can be written down describing $(dL_z/dt)^{\rm tot}$ and $(dQ/dt)^{\rm tot}$.

Because we focus here on equatorial orbits, only terms with $k = 0$ contribute, so we continue to omit the $k$ index.  The infinite sums in Eqs.\ (\ref{eq:dEdttot1})--(\ref{eq:dEdttot4}) must be truncated at finite values.  We truncate the sum over $l$ at $l = l_{\rm max}$ when the next term changes the accumulated total by a fractional amount $\epsilon_l$ for three terms in a row.  In other words, when $l = l_{\rm max}$, the contributions from $l_{\rm max} - 2$, $l_{\rm max} - 1$, and $l_{\rm max}$ change $\dot E_m$ by a fractional amount less than $\epsilon_l$.  We likewise truncate the sum over $m$ at values $m_{\rm neg-max}$ and $m_{\rm pos-max}$ when we find three terms in a row (in both directions) that change $\dot E_k$ by a fractional amount less than $\epsilon_m$.  If we include non-zero $k$ terms, we truncate at $k_{\rm neg-max}$ and $k_{\rm pos-max}$ when we find three consecutive terms that change $\dot E_n$ by a fractional amount less than $\epsilon_k$.  Finally, we truncate the sum over $n$ at $n_{\rm max}$ when we find three terms in a row that change $dE/dt$ by a fraction less than $\epsilon_n$.  As discussed in Sec.\ \ref{sec:amps_and_evolve}, terms with $n < 0$ can be found using the symmetry (\ref{eq:Zsymmetry}).

For this paper, we have used $\epsilon_l = 10^{-7}$, $\epsilon_m = 10^{-6}$, $\epsilon_n = 10^{-5}$.  We do not claim that these values suffice for all purposes (indeed, we are quite sure that more stringent choices must be made for gravitational-wave data analysis pipelines), but we are confident that with these choices, we can reliably describe how these fluxes behave.

\begin{figure*}[ht]
\includegraphics[width=0.48\textwidth]{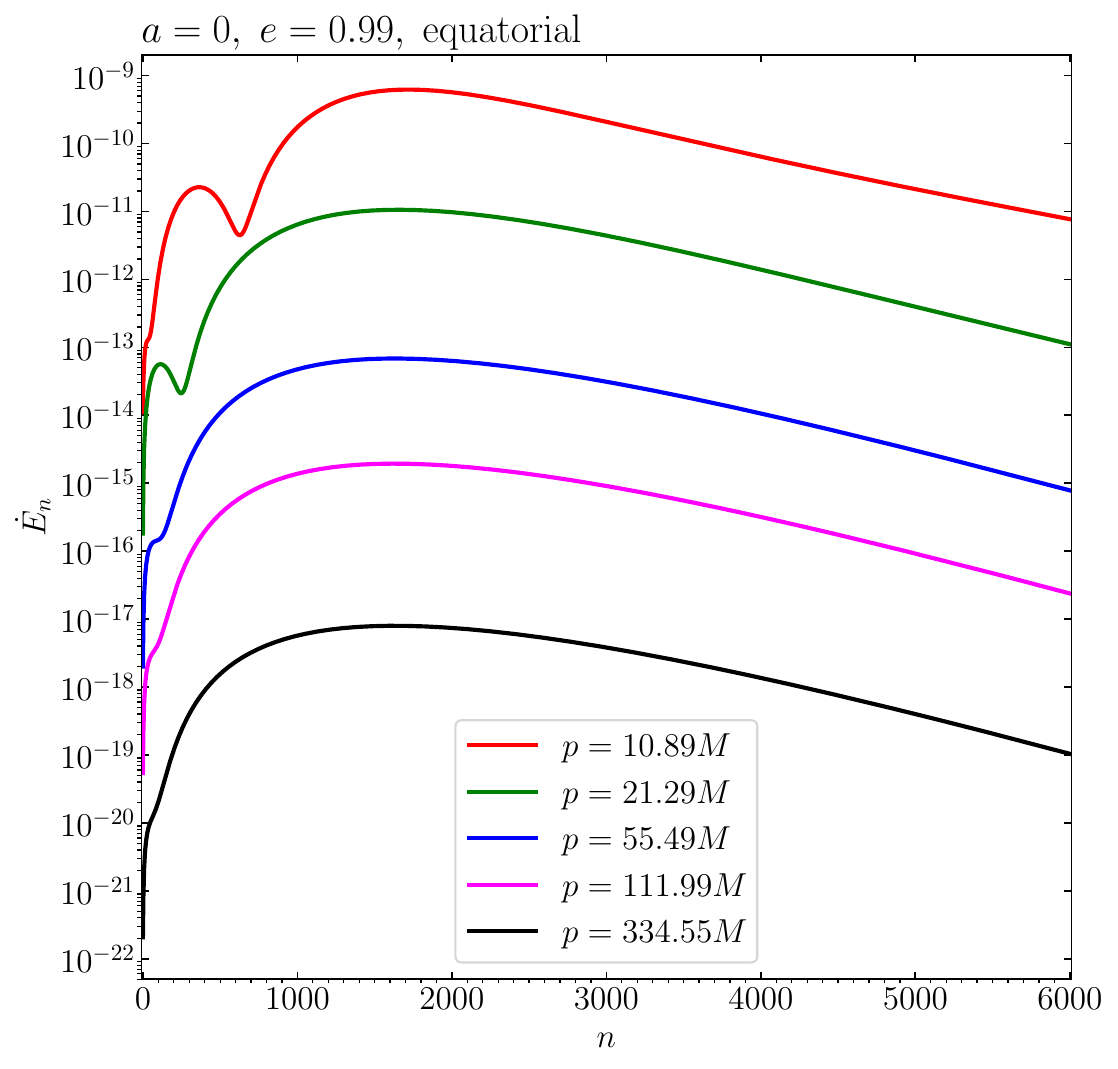}
\includegraphics[width=0.48\textwidth]{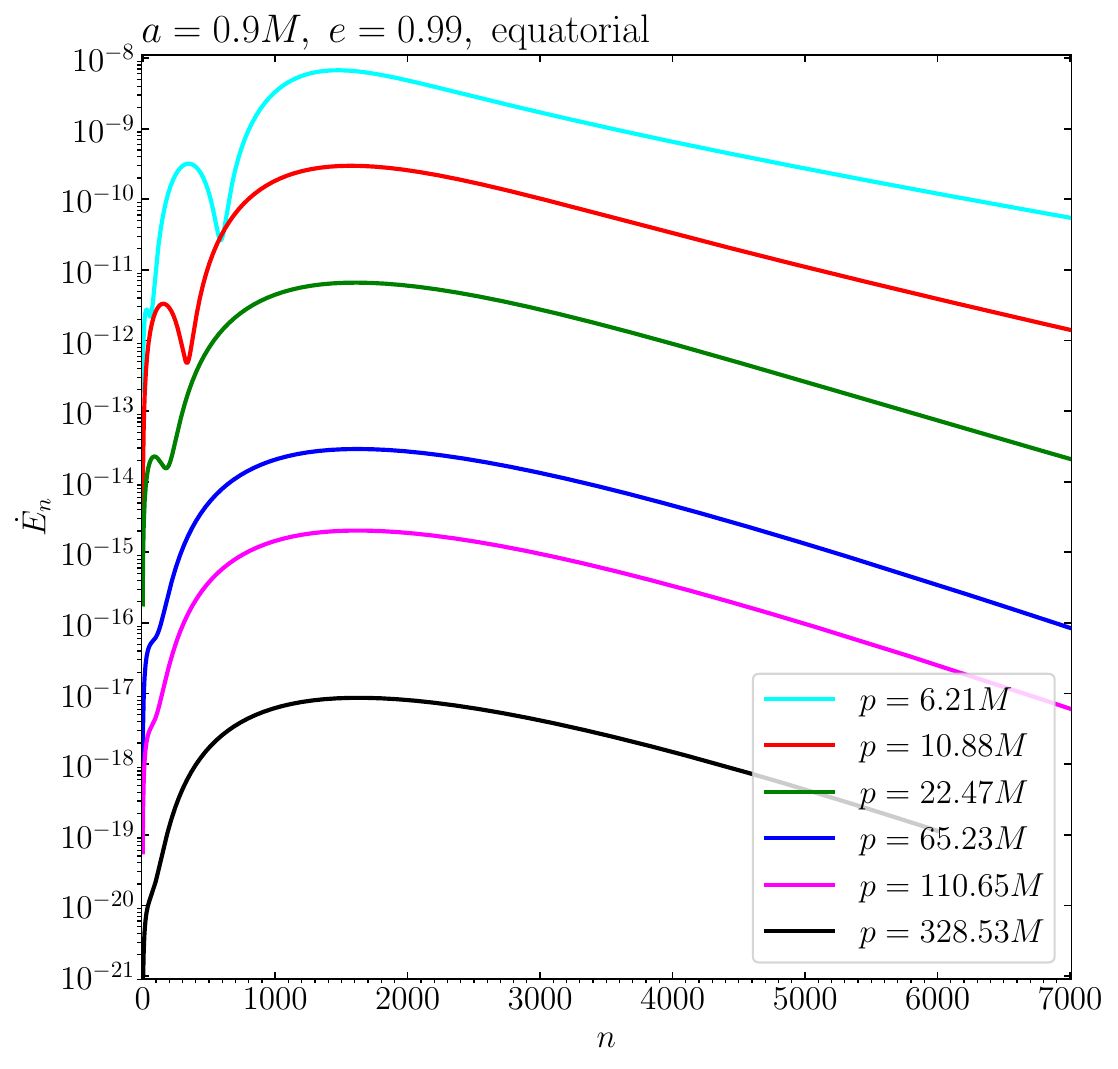}
\caption{The spectrum of energy radiated by gravitational waves in the radial harmonic for a sampling of Schwarzschild (left) and $a = 0.9M$ Kerr orbits, all for equatorial configurations with $e = 0.99$.  All data were computed using the subdivided integral method described in Sec.\ \ref{sec:be_cool_or_be_cast_out}.  The labels show the value of $p$ for these orbits, with the listing order in the inset (from top to bottom) corresponding to the ordering of the spectra in the plot.  The key point to emphasize is that many thousands of harmonics were computed from all these orbits: in each point in the curve, we typically go to $l$ of order several tens (up to $l_{\rm max} = 39$ for the strongest field example); the range of $m$ is roughly from $-l_{\rm max}$ to $l_{\rm max}$.} 
\label{fig:nspectra}
\end{figure*}

\subsection{Energy and angular momentum fluxes\\ versus weak-field formulas}
\label{sec:PMcompare}

In many astrophysical studies, it is common to use weak-field formulas describing the rate of change of an orbit's energy and angular momentum.  Originally derived by Peters and Mathews \cite{Peters:1963ux, Peters:1964zz} (P\&M), these results are derived by treating the binary's orbital kinematics using Newtonian gravity, and computing the backreaction of gravitational waves via the leading-order quadrupole formula.  Considering an extreme mass ratio binary, the resulting weak-field formulas are
\begin{align}
    \left(\frac{dE}{dt}\right)^{\rm P\&M} &= \frac{32}{5}\left(\frac{M}{p}\right)^5 (1 - e^2)^{3/2}f(e)\;,
    \label{eq:dEdtPM}\\
    \left(\frac{dL_z}{dt}\right)^{\rm P\&M} &= \frac{32}{5}\left(\frac{M}{p}\right)^{7/2}(1 - e^2)^{3/2}g(e)\;,
    \label{eq:dLzdtPM}
\end{align}
where 
\begin{equation}
    f(e) = 1 + \frac{73}{24}e^2 + \frac{37}{96}e^4\;,\quad g(e) = 1 + \frac{7e^2}{8}\;.
\end{equation}
Equations (\ref{eq:dEdtPM}) and (\ref{eq:dLzdtPM}) are often written using the semi-major axis of a Keplerian binary; to facilitate comparing with BHPT, we convert to the $(p, e)$ parameterization using $a_{\rm semi} = p/(1 - e^2)$.  The formula for $dE/dt$ is normalized by a factor $(\mu/M)^2$, and $dL_z/dt$ by a factor $\mu^2/M$ in order to facilitate the BHPT comparison.

These weak-field formulas are typically used to approximate an adiabatic evolution of a binary due to gravitational wave driven backreaction.  We assume that orbital kinematics allows us to map between the integrals of motion $(E, L_z)$ and the orbital geometry $(p, e)$.  We then begin with some initial condition $(p_0, e_0)$, map to the initial integrals of motion $(E_0, L_{z0})$, and compute how the orbital energy and angular momentum evolve using Eqs.\ (\ref{eq:dEdtPM}) and (\ref{eq:dLzdtPM}).  At each step in the evolution, our understanding of the kinematics lets us map back to the orbital geometry.  We thus approximate the gravitational wave driven inspiral as an evolution through a sequence of bound orbits.  The adiabatic prescription allows us to construct the sequence of orbits $(p(t), e(t))$ that the system follows in this evolution.

\begin{figure*}[ht]
\includegraphics[width=0.321\textwidth]{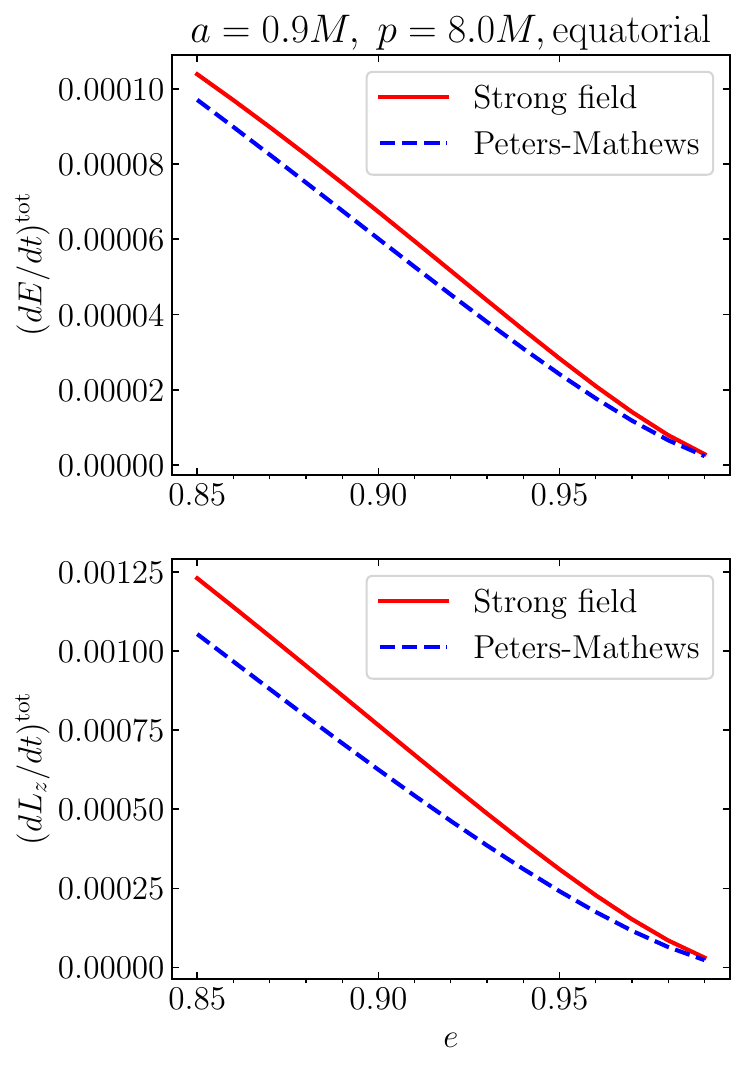}
\includegraphics[width=0.30\textwidth]{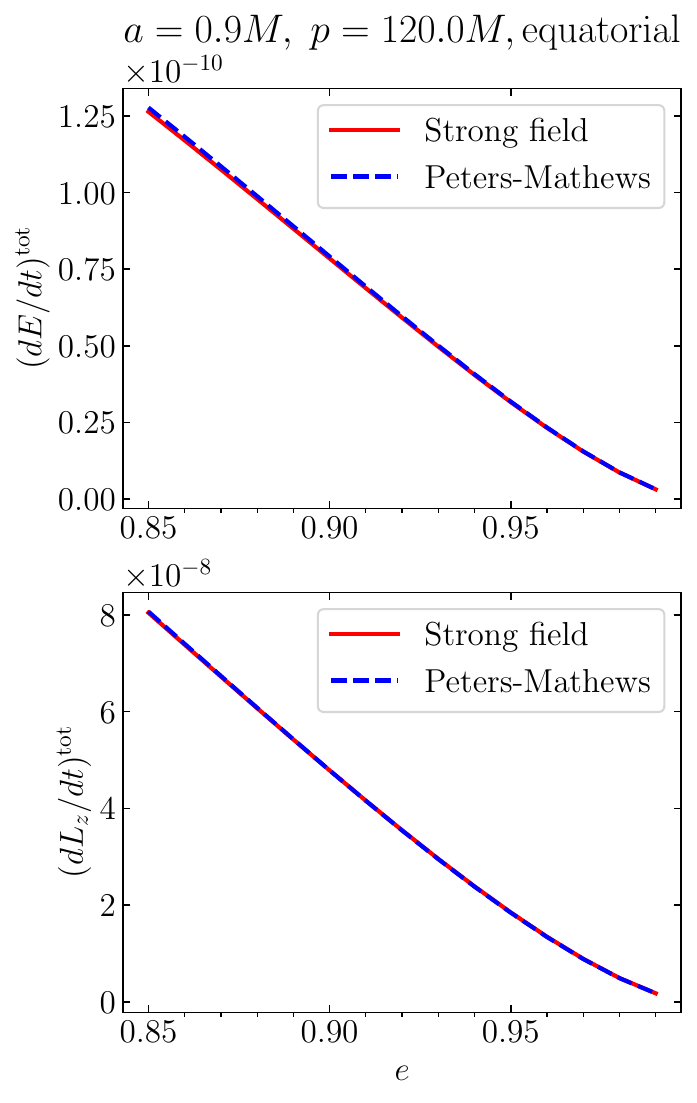}
\includegraphics[width=0.32\textwidth]{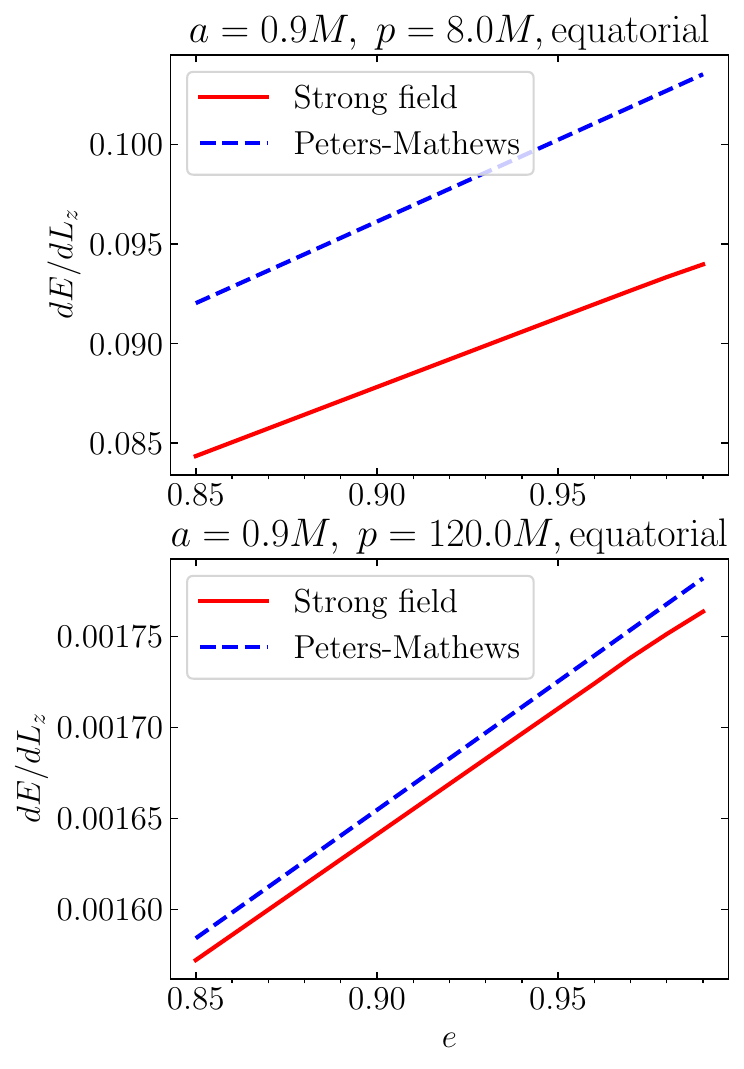}
\caption{Comparison of total fluxes $dE/dt$ and $dL_z/dt$ computed using black hole perturbation theory (BHPT), versus those computed using the weak-field Peters \& Mathews \cite{Peters:1963ux, Peters:1964zz} (P\&M) formulas.  All panels show results for prograde equatorial orbits of a Kerr black hole with spin $a = 0.9M$.  Left-hand panels show these fluxes versus eccentricity for $p = 8M$; center panels show the same data for $p = 120M$.  The right-hand panels show the ratio $dE/dL_z \equiv (dE/dt)/(dL_z/dt)$ corresponding to the data in these panels.  In all panels, the solid (red) curve shows data computed using BHPT; the dashed (blue) curve shows data computed using P\&M.  The difference between the two approaches appears to narrow as $e \to 1$ (consistent with the fluxes apparently vanishing in this limit according to the P\&M formulas).  Notice, however, that the difference in the ratio $dE/dL_z$ computed using the two methods does not narrow as $e$ grows (indeed, it widens slightly).  As discussed in more detail in the text, this flags a systematic error which will accumulate secularly over an inspiral when using the P\&M formulas.}
\label{fig:sf_vs_wf}
\end{figure*}

BHPT lets us assess how well the P\&M formulas describe a binary system.  Figure \ref{fig:sf_vs_wf} compares the fluxes we compute using BHPT with fluxes found using P\&M formulas.  We focus on prograde equatorial orbits of a black hole with $a = 0.9M$, very large eccentricity (covering the range $0.85 \le e \le 0.99$), and examine two orbits: a strong-field case ($p = 8M$), and a relatively weak-field case ($p = 120M$).

Not surprisingly, the comparison shows that the agreement with weak-field formulas is good for the weak-field case --- the two data sets lie nearly on top of each other, differing by less than a percent.  The data differ much more for the strong-field case, by roughly $12\%$.  Interestingly, a substantial and important difference persists across a wide range of the parameter space: the right-most panels of Fig.\ \ref{fig:sf_vs_wf} show the ratio $dE/dL_z \equiv (dE/dt)/(dL_z/dt)$.  This ratio determines the sequence of orbit integrals $(E(t), L_z(t))$ that the system follows on its inspiral, and thus determines the sequence $(p(t), e(t))$ that this follows.  Because the P\&M prediction of the ratio $dE/dL_z$ systematically differs from what is found using BHPT, an evolution determined using the P\&M formulas is guaranteed to yield the incorrect sequence $(p(t), e(t))$.

This finding merits more study and analysis.  In particular, this is not an artifact of the high-eccentricity limit that is our focus here, but appears to hold across orbital parameter space.  A detailed study of this behavior, accompanied by the presentation of easy-to-use tools to ``upgrade'' from the P\&M formulas, is in preparation \cite{Becker_inprep}.

\subsection{Convergence of total radiated \\ energy and angular momentum per orbit}
\label{sec:totalperorbit}

The weak-field formulas (\ref{eq:dEdtPM}) and (\ref{eq:dLzdtPM}) vanish as $e \to 1$; at least graphically, the strong-field numerical data shows the same trend.  These quantities represent the orbit-averaged rates of change of an orbit's energy and angular momentum, so the products
\begin{equation}
    \Delta E^\star \equiv \left(\frac{dE}{dt}\right)^\star T_r\;,\quad
    \Delta L_z^\star \equiv \left(\frac{dL_z}{dt}\right)^\star T_r\;.
\end{equation}
are the total change in those orbit integrals over a single orbit cycle.  (The superscripted ``$\star$'' can stand for $\infty$ or ${\rm H}$, depending on whether one focuses on fluxes to infinity or down the horizon; or it can stand for ``tot'' to denote the sum of these contributions.)  Since the radial period $T_r \to \infty$ as $e \to 1$, the behavior of $\Delta E^\star$ and $\Delta L_z^\star$ as $e$ approaches unity can help us to anchor backreaction in this limit.

\begin{figure}[h]
\includegraphics[width=0.457\textwidth]{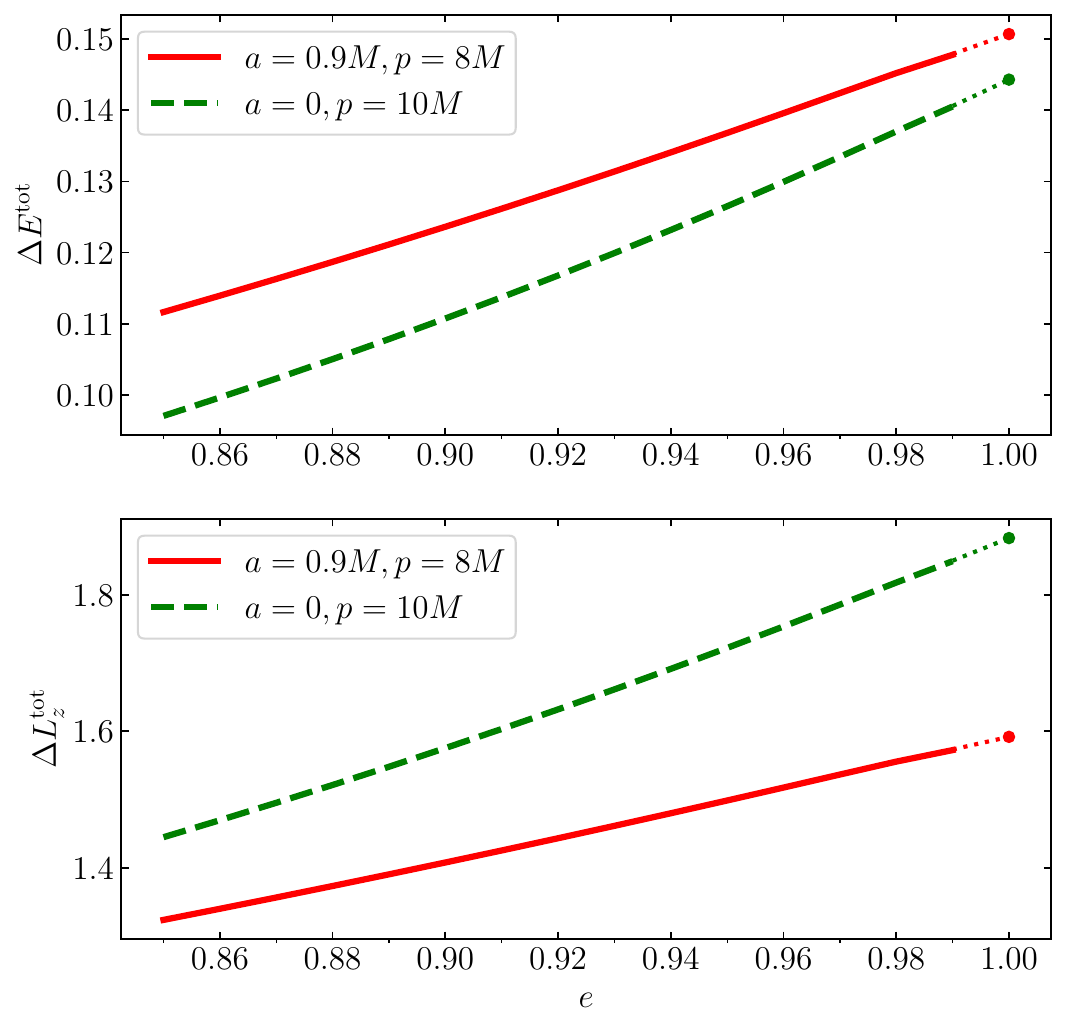}
\caption{The total radiated energy $\Delta E^{\rm tot}$ (top panel) and angular momentum $\Delta L^{\rm tot}_z$ (bottom) per orbit versus $e$.  The solid (red) curve is for $p = 8M$ orbits of a Kerr black hole, $a = 0.9M$; the dashed (green) curve is for $p = 10M$ orbits of a Schwarzschild black hole.  These data include both radiation to infinity and down the horizon.  By fitting a polynomial in $e$ and extrapolating, we are able to estimate the total radiated energy and angular momentum beyond the range of data we computed directly.  The extrapolation is indicated by the dotted portion, with a large dot at $e = 1$.  As discussed in the text, the Schwarzschild data agree very well with past work examining this quantity on parabolic orbits.}
\label{fig:deltaE_deltaL}
\end{figure}

Figure \ref{fig:deltaE_deltaL} shows $\Delta E^{\rm tot}$ and $\Delta L_z^{\rm tot}$ as a function of eccentricity, showing trends for Kerr ($a = 0.9M$) orbits at $p = 8M$, and for Schwarzschild orbits at $p = 10M$.  To produce data for this plot, we use the same convergence criterion as described in Sec.\ \ref{sec:convergedfluxes}, so we expect these results to have a fractional error of about $10^{-5}$.

The data shown in Fig.\ \ref{fig:deltaE_deltaL} show that $\Delta E^{\rm tot}$ and $\Delta L_z^{\rm tot}$ are well behaved as we approach the parabolic orbit limit, $e \to 1$.  To estimate their values in this limit, we fit a low-order (cubic) polynomial in $e$ to these data over the range $0.85 \le e \le 0.99$, and then extrapolate to $e = 1$.  We find
\begin{align}
    (a = 0, p = 10M):\; &\Delta E^{\rm tot} = 0.144288\mu^2/M\;,
    \label{eq:DeltaEtot_1}    \\
    &\Delta L_z^{\rm tot} = 1.88308\mu^2\;;
    \label{eq:DeltaLtot_1}    \\
    (a = 0.9M, p = 8M):\; &\Delta E^{\rm tot} = 0.150679\mu^2/M\;,
    \label{eq:DeltaEtot_2}    \\
    &\Delta L_z^{\rm tot} = 1.59189\mu^2\;.
    \label{eq:DeltaLtot_2}
\end{align}
To clarify how these quantities scale with a binary's masses, we have restored mass factors which had been absorbed into various flux formulas; cf.\ discussion following Eq.\ (\ref{eq:dQdthrz}).  Bearing in mind that $T_r$ scales with the large black hole mass $M$, we find the mass scalings shown in Eqs.\ (\ref{eq:DeltaEtot_1})--(\ref{eq:DeltaLtot_2}).

Repeating this exercise but including only the flux at infinity contribution to these quantities, we find
\begin{align}
    (a = 0, p = 10M):\; &\Delta E^\infty = 0.142446\mu^2/M\;,
    \\
    &\Delta L_z^\infty = 1.86678\mu^2\;;
    \\
    (a = 0.9M, p = 8M):\; &\Delta E^\infty = 0.151918\mu^2/M\;,
    \\
    &\Delta L_z^\infty = 1.60382\mu^2\;.
\end{align}
The value of $\Delta E^\infty$ we find for Schwarzschild $p = 10M$ agrees very well with the value reported for this orbit by Warburton \cite{Warburton:2025ymy} using a code specialized for hyperbolic and parabolic orbits.  Our value is smaller than Warburton's by about $0.015\%$; the agreement would presumably improve if we use a more stringent value of $\epsilon_n$, and perhaps include data in our set for larger $e$ (e.g., $e = 0.995$ or $e = 0.997$).

Notice that $\Delta E^{\rm tot} > \Delta E^\infty$ and $\Delta L_z^{\rm tot} > \Delta L_z^\infty$ for our Schwarzschild example, but we find $\Delta E^{\rm tot} < \Delta E^\infty$ and $\Delta L_z^{\rm tot} < \Delta L_z^\infty$ for Kerr.  This indicates that the energy and angular momentum absorbed by the black hole is negative for Kerr, which is consistent with the expectation that many modes which contribute to these terms are superradiant for this rapid black hole spin.

\section{Discussion}
\label{sec:discussion}

In this analysis, we have demonstrated that straightforward modifications allow frequency-domain BHPT-based codes to significantly extend the parameter-space coverage of small-ratio binary systems.  We have found that modeling binary systems with eccentricity $e = 0.99$ is not difficult; it appears that coverage to $e = 0.997$ can be achieved without too much difficulty.  We have pushed our calculation to $e = 0.999$, though numerical noise degrades precision at this eccentricity.  The value $e \simeq 0.997$ may be a practical limit for producing precise models using frequency-domain BHPT.

These results may be useful for astrophysical studies of the formation of eccentric binary systems, particularly for analyses which show systems evolving, even transiently, through regimes in which $1 - e$ is very small (e.g., Refs.\ \cite{Qunbar:2023vys, Rom:2024nso, Xuan:2025uvl, Mancieri:2025cmx}).  We have found that the classic Peters and Mathews \cite{Peters:1963ux, Peters:1964zz} results describing the leading-order evolution of binary systems due to gravitational wave backreaction systematically miscalculate $dE/dL_z$ even for relatively weak-field systems.  This error will secularly accumulate, skewing the trajectories that an inspiraling system follows in $(E, L_z)$ parameter space, and thus systematically mispredicting the distribution of binaries in $(p, e)$.  Perhaps this effect is small, but it should be examined and quantified; improving the treatment of strong gravity physics is known to have a highly non-trivial impact on models of binary formation in this regime \cite{Alush_inprep}.  Work in progress \cite{Becker_inprep} aims to produce easy-to-use tools that would allow BHPT data to be used in such analyses; results from this study will be incorporated into that work to allow those tools to cover the high-eccentricity end of parameter space.

Largely to simplify the discussion and to reduce computation time, we have focused on equatorial orbits in all the cases shown here, but the code and techniques we have developed handle generic (inclined and eccentric) orbits without problem.  We have in fact computed several examples of high-eccentricity, highly inclined orbits as tests of the code and as input to other analyses which will be presented elsewhere \cite{Alush_inprep}.  Computing generic cases tends to be quite computationally expensive.  Generating BHPT solutions for generic orbits requires evaluating a 2-D source integral (we must also integrate the $\lambda_\theta$ parameter over the source's polar motion), and the modes depend on an additional harmonic index $k$, describing harmonics of the polar orbit frequency.  Generic sources also tend to generate large data files, thanks to the 4-index harmonic structure of their modes (compared to the 3-index solution describing equatorial orbits).  But no issue of principle prevents detailed study of such cases.

It may be useful to combine studies using the techniques we describe in this paper with those using special codes which treat $e = 1$ separately, such as was done in Ref.\ \cite{Warburton:2025ymy} for Schwarzschild.  It may also be useful to explore in more detail whether the integration by parts method can yield better accuracy.  We suspect that for some mode number, we will simply hit a noise floor associated with the limits of double-precision numerical arithmetic; the limiting case we included ($n \gtrsim 120000$ for $e = 0.999$) is perhaps already affected by that limit.  Further analysis may clarify whether going further is possible or practical with codes like the one we have used here.

\acknowledgements

We thank Yael Alush, Smadar Naoz, Barak Rom, Nicholas Stone, and Zeyuan Xuan for many discussions that motivated much of this analysis (including feedback from Xuan on a draft of this paper), Christopher Whittal whose Ph.D.\ viva provided technical inspiration for some of the techniques we tested to tame the high-eccentricity limit, Devin Becker for many useful discussions about the behavior of these data across parameter space, and Niels Warburton for very useful feedback and comments on a draft of this paper.  This work was supported by NSF Grant PHY--2409644 and the MIT Undergraduate Research Opportunities Program (UROP).  Computations were done using the MIT {\tt engaging} cluster, which is supported by the MIT Office of Research Computing and Data, as well resources provided by {\tt subMIT} at MIT Physics.  {\tt SubMIT} facilitates access to the resources of the Open Science Grid Consortium \cite{osg1, osg2, osg3, osg4}, which is supported by National Science Foundation Awards 2030508 and 2323298.

\appendix

\section{Lengthy formulas}
\label{app:lengthy}

In this appendix, we present lengthy formulas that are needed in various places, but whose details are not immediately needed in this paper's main body text.  First, the factor $\mathcal{L}_{mkn}$ which appears in the rates of change of $Q$, Eqs.\ (\ref{eq:dQdtinf}) and (\ref{eq:dQdthrz}), is given by
\begin{equation}
{\mathcal L}_{mkn} = m\langle\cot^2\theta\rangle L_z -a^2\omega_{mkn}\langle\cos^2\theta\rangle E\;.
\label{eq:Lmkndef}
\end{equation}
The terms $\langle\cot^2\theta\rangle$ and $\langle\cos^2\theta\rangle$ in Eq.\ (\ref{eq:Lmkndef}) mean $\cot^2\theta$ and $\cos^2\theta$ evaluated at the $\theta$ coordinate along the orbit, and then averaged using Eq.\ (\ref{eq:rthetaaveraging}).  The factor $\alpha_{lmkn}$ which appears in the down-horizon rates of change (\ref{eq:dEdthrz})--(\ref{eq:dQdthrz}) is given by
\begin{widetext}
\begin{equation}
\alpha_{lmkn} = \frac{256(2Mr_+)^5(\omega_{mkn} - m\Omega_{\rm H})[(\omega_{mkn} - m\Omega_{\rm H})^2 + 4\epsilon^2][(\omega_{mkn} - m\Omega_{\rm H})^2 + 16\epsilon^2]\omega_{mkn}^3}{|C_{lmkn}|^2}\;,
\label{eq:alphadef}
\end{equation}
where
\begin{align}
\Omega_{\rm H} &= \frac{a}{2Mr_{+}}\;,
\label{eq:OmegaH}\\
\epsilon &= \frac{\sqrt{M^2 - a^2}}{4Mr_+}\;,
\label{eq:epsilon}\\
|C_{lmkn}|^2 &= [(\lambda_{lmkn}^2 + 2)^2 + 4am\omega_{mkn} - 4a^2\omega_{mkn}^2](\lambda_{lmkn}^2 + 36am\omega_{mkn} - 36a^2\omega_{mkn}^2)
\nonumber\\
&+ (2\lambda_{lmkn} + 3)(96a^2\omega_{mkn}^2 - 48am\omega_{mkn}) + 144\omega_{mkn}^2(M^2 - a^2)\;.
\label{eq:Clmknsq}
\end{align}
The number $\lambda_{lmkn}$ is related to the eigenvalue of the spheroidal harmonic $S_{lm\omega}(\theta)$; see Refs.\ \cite{Drasco:2005kz, OSullivan:2014ywd} for a precise definition.

The factors $A_{abi}$ which appear in the source integral function $I^{\infty, {\rm H}}_{lm\omega}(r,\theta)$, Eq.\ (\ref{eq:source_integrand}), are given by
\begin{align}
    A_{nn0} &= -\frac{2 \rho^{-3}\bar \rho^{-1}C_{nn}}{\Delta^2}\left(L_1^\dagger L_2^\dagger S - 2ia\rho L_2^\dagger S \sin \theta\right)\;,
    \\
    A_{n \bar m 0} &= \frac{2\sqrt{2}\rho^{-3}C_{n\bar m}}{\Delta} \left[\left(\frac{iK}{\Delta} + \rho + \bar \rho\right)L_2^\dagger S +\left(\frac{K}{\Delta}\right)a\sin\theta S(\rho - \bar\rho)\right]\;,
    \\
    A_{\bar m \bar m 0} &= S \rho^{-3}\bar \rho C_{\bar m \bar m} \left[\left(\frac{K}{\Delta}\right)^2 - 2i\rho\frac{K}{\Delta} + i \partial_r\left(\frac{K}{\Delta}\right) \right]\;,
    \\
    A_{n \bar m 1} &= \frac{2 \sqrt{2}\rho^{-3}C_{n \bar m}}{\Delta}\left[L_2^\dagger S + ia (\bar\rho - \rho)S \sin \theta\right]\;,
    \\
    A_{\bar m \bar m 1} &= -2 S \rho^{-3} \bar \rho\  C_{\bar m \bar m} \left(\rho + \frac{iK}{\Delta}\right)\;,
    \\
    A_{\bar m \bar m 2} &= -S \rho^{-3} \bar \rho\  C_{\bar m \bar m}\;.
\end{align}
In these equations, $\rho = 1/(r - ia\cos\theta)$, $\bar\rho = 1/(r + ia\cos\theta)$, and $S = S_{lm\omega}(\theta)$.  Precise definitions of the angular derivative operators $L^\dag_1$ and $L^\dag_2$ can be found in Refs.\ \cite{Hughes:1999bq, Drasco:2005kz}.  The various $C_{ab}$ in these terms gather factors that group together when we evaluate projections of the Newman-Penrose legs $n^\alpha$, $\bar m^\alpha$ with the stress-energy tensor $T_{\alpha\beta}$:
\begin{align}
    C_{nn} &= \frac{\mu}{4 \Sigma^2(dt/d\lambda)}\left[E(r^2+a^2)-aL_z+\frac{dr}{d\lambda}\right]^2\;,
    \label{eq:Cnn}\\
    C_{\bar m\bar m} &= \frac{\mu \rho^2}{2(dt/d\lambda)}\left[i\left(a\sin\theta E - \csc\theta L_z\right) + \frac{d\theta}{d\lambda}\right]^2\;,
    \label{eq:Cmbarmbar}\\
    C_{n\bar m} &= -\frac{\mu \rho}{2\sqrt{2}\Sigma(dt/d\lambda)}\left[E(r^2 + a^2) - aL_z+\frac{dr}{d\lambda}\right] \left[i\left(a\cos\theta E - \csc\theta L_z\right) + \frac{d \theta}{d \lambda}\right]\;.
    \label{eq:Cnmbar}
\end{align}

\end{widetext}

\bibliography{sahughes_refs}

\end{document}